\documentclass[letter,12pt]{article}
\pdfoutput=1 

\usepackage{jheppub} 

\usepackage[utf8]{inputenc}

\usepackage{epsfig}
\usepackage{amssymb}
\usepackage{amsfonts}
\usepackage{amsmath}
\usepackage{titlesec}
\usepackage{yfonts}
\usepackage{booktabs}
\usepackage{mathrsfs}
\newcommand{\qbinom}[2]{\genfrac{[}{]}{0pt}{}{#1}{#2}_p}

\usepackage{adjustbox}

\usepackage{arydshln}

\usepackage{ dsfont }

\usepackage[english]{babel}
\usepackage[T1,T2A]{fontenc} 

\usepackage[utf8]{inputenc}
\usepackage{amsmath}
\usepackage{calc}
\usepackage{bm}

\usepackage{accents}

\newcommand{\scalar}[1]{ {\langle} {#1}  {\rangle} }

\def\IQ{{\mathbb Q}}

\def\T{{\mathcal T}}

\def\coset{{\mathscr C}}

\def\NC{{\mathcal N}}

\def\Mg{\Sp(2g,\Z)}

\def\H{{\mathcal H}}
\def\D{{\mathcal D}}

\def\C{{\mathcal C}}

\def\1{\mathbb{I}}

\def\M{ M}
\def\N{{\mathcal N}}

\def\B{{\mathcal B}}
\def\S{{\mathcal S}}

\def\Sa{{S}_h}

\def\G{ {\mathcal G}}
\def\FF{ {\mathcal F}}

\newcommand{\Z}{\ensuremath{\mathbb Z}}
\newcommand{\R}{\ensuremath{\mathbb R}}

\newcommand{\kket}[1]{\left|\!\left| #1 \right\rangle\!\right\rangle}

\usepackage{amsmath}

\usepackage{ upgreek }
\usepackage{physics}

\def\bea{\begin{eqnarray}}
\def\eea{\end{eqnarray}}

\newcommand{\Sp}{\mathrm{Sp}}
\newcommand{\SL}{\mathrm{SL}}
\newcommand{\GL}{\mathrm{GL}}
\newcommand{\SO}{\mathrm{SO}}
\newcommand{\Or}{\mathrm{O}}
\newcommand{\Un}{\mathrm{U}}

\mathchardef\mhyphen="2D

\title{Mass formula for topological boundary conditions from TQFT gravity}

\author{Anatoly Dymarsky${}^{a,b}$ and Alfred Shapere${}^b$} 
\affiliation{${}^a$ School of Natural Sciences, \\  Institute for Advanced Study, \\ 1 Einstein drive, Princeton, NJ,  08540, USA\\}

\affiliation{${}^b$ Department of Physics and Astronomy, \\ University of Kentucky,\\ 505 Rose Street, Lexington, KY,  40506, USA\\}
\emailAdd{a.dymarsky@uky.edu}
\emailAdd{shapere@g.uky.edu}

\abstract{
Mass formulas  
evaluate the total weighted count of a given class of algebraic structures, such as lattices or codes. 
We show that 3d TQFTs provide a generalization of 
this concept:  the total weighted count of topological boundary conditions  is given by the TQFT partition function averaged over all closed 3d manifolds.  This weighted count, which we call the mass, can be interpreted as the renormalized partition function of TQFT gravity.   For Abelian TQFTs, the mass formula for topological boundary conditions reduces to the mass formula for particular families of codes. Focusing on the Abelian  case, we show how to evaluate the mass for any bosonic theory and consider many explicit examples. 
We then discuss  the non-Abelian generalization and compute the mass for $n+\bar n$ copies of the Ising modular tensor category.  Finally, we generalize the construction to five dimensions and compute the mass for Abelian 
$2$-form Chern-Simons theories.

}

\begin{document} 
\maketitle
\flushbottom

\section{Introduction}
In mathematics, mass formulas evaluate the total weighted count of codes, lattices, or other algebraic structures.\footnote{A more general notion underlying mass formula is homotopy cardinality
\cite{baez2001finite}.}  In particular, in coding theory  mass formulas evaluate the total number of inequivalent codes of a given type.  For example, for Type II codes of length $n$,\footnote{These are the doubly-even self-dual  linear binary codes. See the definition in section \ref{dec} below.} the mass formula takes the following form \cite{MACWILLIAMS1972153,PLESS1975313}
\bea
\label{masstypeII}
\sum_{[\C]} {1\over \left|{\rm Aut}(\C)\right|} = {\NC\over |G|}, \qquad \NC(n)=\prod_{i=0}^{n/2-2}(2^i+1).
\eea
The sum here goes over equivalence classes of codes $[\C]$. The expression on the RHS is known as the ``mass.'' It is  the total number of codes
$\NC$ divided by the order  $|G|=n!$  of the code equivalence group,  which in this case is the  group of permutations.  
After multiplying both sides by $|G|$,
the mass formula can be rewritten as a sum over all codes $\C$,  in which case it  simply counts the total number of codes, $\sum_\C 1 =\N$. 

Fast forward to 3d TQFTs (of the unitary modular tensor category type). Our main interest in this paper will be to count topological boundary conditions (TBCs) \cite{kapustin2011topological,Kapustin:2010if,kaidi2022higher}, which we will denote by $\C$. In \cite{Dymarsky:2024frx} we have shown that TQFT gravity -- a model of topological gravity defined as a  topological field theory $\mathcal T$ summed over all possible 3d topologies
with fixed boundary -- 
is holographically dual to a weighted ensemble of boundary theories indexed by $\C$. For an Abelian theory $\T$, 
the sum over closed manifolds  yields the following mass formula for the total number of TBCs,
\bea
\label{mass}
\NC(\T)=\sum_\C 1= \lim_{g\rightarrow \infty} {\D^g\over |\Mg|} \sum_{\gamma \in \Mg} {}^g\langle 0|U_\gamma|0\rangle^g.
\eea
We will explain this formula and review its derivation of this formula in section \ref{derivation}, and consider
its generalization to non-Abelian $\mathcal T$  in section \ref{nonAbelian}. 
The sum on the RHS is the TQFT partition function averaged over  all possible  closed connected 3d manifolds, represented by their Heegaard splittings (and multiplied by a power of the total quantum dimension $\D$ of $\T$). We  interpret it as the renormalized partition function of the TQFT gravity defined by $\T$.

When $\T$ is Abelian, its TBCs $\C$ can be identified with classical even self-dual codes \cite{Barbar:2023ncl}, and 
\eqref{mass} reduces to the conventional mass formulas for various  families of self-dual codes. In particular, \eqref{masstypeII} will follow from the mass formula for $\Un(1)_2^n$ Chern-Simons theory, as  discussed in section \ref{dec} below.

One of the  goals of this paper will be to show how the sum over all closed 3d manifolds in \eqref{mass} can be evaluated for any Abelian bosonic theory $\T$.  A well-established conventional approach is to parameterize the topologies of closed 3d manifolds in terms of links in $\mathbb{S}^3$, with the corresponding Abelian TQFT partition function given by a Gauss sum \cite{deloup1999linking,deloup2001abelian,Guadagnini:2013sb,kaidi2022higher,Kim:2024shg,Nicosanti:2025xwu}. Instead we parametrize closed 3d manifolds in terms of Heegaard splittings. One of the tools we use along the way is a particular basis in the TQFT Hilbert space that draws on the connection with quantum stabilizer codes and simplifies the action of the mapping class group.

Our paper extends ideas of \cite{kaidi2022higher} that connect a given TBC with the partition function of the underlying theory $\T$ on certain lens spaces. 
For Abelian theories,  it was shown in \cite{kaidi2022higher} that TBCs exist if and only if all such partition functions are real, i.e.~whenever all higher central charges vanish. As a spinoff application, we will show how this result follows from our formalism. 

Beyond the Abelian case, the mass formula can be defined for any  theory $\T$.  In the non-Abelian case the weight factors are non-trivial and the mass is not simply
the total count of TBCs.
 We discuss this in  section \ref{nonAbelian},  where we also  evaluate the mass for $n+\bar n$ copies of the Ising category ($n$ chiral and $\bar n$ anti-chiral theories). 

The holographic duality involving TQFT gravity can be extended to other dimensions \cite{Barbar:2025krh}. We leverage that to evaluate the number of maximal non-anomalous subgroups of 2-form symmetry in five-dimensional $2$-form Chern-Simons theory by summing it over closed 5d geometries. 

The paper is organized as follows. In section \ref{prelim} we focus on Abelian bosonic 3d TQFTs and derive the mass formula \eqref{mass}, while also introducing the tools necessary to evaluate the sum over the Heegaard splittings.   We explain how this sum is evaluated in section \ref{sum3d}, where many explicit examples are considered.  The non-Abelian case is considered in section \ref{nonAbelian}, while the generalization to five-dimensional Abelian theories is discussed in section \ref{sec:5d}. We conclude with a summary in section \ref{summary}.

%

\label{sec:intro}

\section{Derivation of the mass formula}
\label{prelim}

\subsection{Preliminaries:  Abelian TQFTs}
We consider an  Abelian Dijkgraaf-Witten  theory \cite{Dijkgraaf:1989pz} (bosonic Abelian Chern-Simons)   $\mathcal T$ specified by an Abelian group $\mathscr{D}$ and a quadratic refinement $q: \mathscr{D}\rightarrow \IQ/\Z$, which defines a scalar product on $\mathscr{D} \times \mathscr{D}$, 
\bea
\label{scalar}
\scalar{c,c'}=2^{-1}(q(c+c')-q(c)-q(c')).
\eea
It  can be represented as an Abelian Chern-Simons theory, with $K$-matrix corresponding to the scalar product \eqref{scalar}; the $K$-matrix is the Gram matrix of  a lattice $\Lambda$ whose discriminant group is $\mathscr{D}=\Lambda^*/\Lambda$.

We are interested in theories admitting  topological boundary conditions, which requires  $c_-=0~{\rm mod}~8$. 
Provided this is satisfied, we assume  the chiral central charge is canceled by adding the necessary number of topologically trivial $(E_8)_1$ theories \cite{kaidi2022higher}. 

Consider such a theory on $\Sigma_g \times \R$, where $\Sigma_g$ is a Riemann surface  of genus $g$. Let $X_3$ be a handlebody ending on $\Sigma_g$. 
We introduce the conventional orthonormal basis vectors for the Hilbert space on $\Sigma_g$  \cite{Belov:2005ze}
\bea
\label{basis}
|\vec{c}\, \rangle \in  \H^g ,\qquad \vec{c}\in  \mathscr{D}^g.
\eea  
Here $|\vec{c}\, \rangle$ is the path integral on $X_3$ with Wilson lines of charges $c_i$ wrapping non-shrinkable cycles of $\Sigma_g$. 
The vacuum state, which is the path integral on $X_3$ without any insertions, will be denoted by $|\vec{0}\rangle\equiv |0\rangle^{g}$. We normalize it so that ${}^g\langle 0|0\rangle^g=1$, i.e.~the path integral on the connected sum of $g$ copies of $\mathbb{S}^2\times\mathbb{S}^1$, obtained by gluing the boundaries of two copies of $X_3$, will be given by ${}^g\langle 0|0\rangle^g\,\D^{g-1}=\D^{g-1}$. Here $\D=|\mathscr{D}|^{1/2}$ is the total quantum dimension of $\T$. 


In this basis the modular data -- the $T$ and $S$ matrices -- have the following form
\bea
T_{cc'}&=&\theta_{c\,}\delta_{cc'},\qquad \quad\quad \theta_c=e^{2\pi i q(c)}\equiv e^{i\pi \scalar{c,c}},\\
S_{cc'}&=&{1\over \D} e^{-2\pi i \scalar{c,c'}}. 
\eea

In an Abelian bosonic TQFT, topological boundary conditions (TBCs)  correspond to Lagrangian subgroups of $\mathscr{D}$ -- non-anomalous maximal subgroups $\C\subset  \mathscr{D}$ \cite{ kapustin2011topological,Kapustin:2010if,kaidi2022higher}. The Lagrangian condition can be reformulated as follows: for any $c\in \C$,
\bea
\label{evenness}
&&\scalar{c,c}=2q(c)=0\, \,{\rm mod}\, \,2,\\ 
&&\scalar{c,c'}=0\, \,{\rm mod}\,\, 1 ,\quad {\rm iff}\, c'\in \C. \label{selfduality}
\eea 
These conditions can be interpreted in  coding theory as evenness and self-duality of the code $\C$  \cite{Barbar:2023ncl};\footnote{
The interpretation of Lagrangian subgroups as codes readily gives rise to the code CFTs of \cite{Barbar:2023ncl,dolan1996conformal,DOLAN1990165,Dymarsky:2020qom, Dymarsky:2020bps,Dymarsky:2021xfc,Henriksson:2022dnu,
Henriksson:2022dml,Buican:2021uyp,Angelinos:2022umf,Furuta:2022ykh,Alam:2023qac, Furuta:2023xwl, Kawabata:2023nlt, Kawabata:2023iss,Kawabata:2023rlt,  Buican:2023ehi,Angelinos:2025mjj}, starting from the SymTFT construction.} various explicit examples will be discussed below.  
We note that when the code is even, it satisfies \eqref{evenness}; self-duality \eqref{selfduality} is then equivalent to $\C$ being a maximal subgroup of $\mathscr{D}$, i.e.~a subgroup of order $\D$.

In the context of codes, an element $c\in \C$ is conventionally called a codeword while the list of all possible codewords $\mathscr{D}$ is called the ``dictionary.''
The language of codes is most natural when the theory is a product of $n$ equal copies $\T^{\otimes n}$, in which case the dictionary (the disciminant group) is $\mathscr{D}^n$ and $ \mathscr{D}$ is conventionally called the alphabet. 

A topological boundary condition $\C$ defines a state 
\bea
\label{Cstate}
|{\cal C}\rangle=\sum_{\vec{c}\in {\C^g}}|\vec{c}\,\rangle \in \H^g,
\eea
which is the  TQFT path integral  on the cylinder $\Sigma_g \times [0,1]$ with the topological boundary condition $\C$ imposed at one end.
This state is invariant under the mapping class group $\Sp(2g,\Z)$ of $\Sigma$.\footnote{Except in section \ref{nonAbelian}, we will generally equate the mapping class group of $\Sigma_g$ with $\Sp(2g,\Z)$,  which is  the quotient of the mapping class group by its Torelli subgroup, the subgroup that acts trivially on 1-homology.  The latter acts trivially in the Abelian case.}  
We normalize it so that $\langle \C|0\rangle^g=1$. 

\subsection{Mass formula from the sum over topologies}
\label{derivation}
The idea behind our approach to the mass formula is as follows \cite{Dymarsky:2024frx}. 
States $|\C\rangle\in \H^g$, defined in terms of topological boundary conditions, are invariant under the action of the mapping class group on $\H^g$,
\bea
U_\gamma |\C\rangle =|\C\rangle,\qquad \gamma\in \Sp(2g,\Z).
\eea 
When the genus $g$ is small, the states $|\C\rangle$ are often linearly dependent, but they become linearly independent for $g\geq \D$. 
Crucially, when the theory is Abelian the states $|\C\rangle$ are believed to span the space of modular invariants in $\H^g$ for any $g$. A rigorous argument to that effect is known in the coding literature as Gleason's theorem \cite{sloane2006,Nebe}.\footnote{To apply Gleason's theorem, one should identify the state $|\C\rangle$ with the code's full enumerator polynomial, see e.g.~\cite{Dymarsky:2025agh} where this identification is explained in detail.  In fact the proof in \cite{sloane2006} is somewhat weaker than necessary; it applies not to states $|\C\rangle$ but to wavefunctions written in terms of characters. It would be interesting to translate the result of  \cite{sloane2006}  into the language of Abelian Dijkgraaf-Witten theory. }

When the genus $g$  is very large, the TBC states $|\C\rangle$ become mutually orthogonal and, since they span the space of modular invariants in $\H^g$,
 the density matrix  
\bea
\label{proj}
\rho=\sum_\C {|\C\rangle \langle \C|\over\langle \C|\C\rangle} \in \H^g \otimes \bar \H^g
\eea
is  a projector onto the modular-invariant subspace inside $\H^g$.  It therefore can be written as a formal average over the mapping class group
\bea
\label{proj2}
\rho={1\over |\Sp(2g,\Z)|}\sum_{\gamma\in \Sp(2g,\Z)}  U_\gamma,\qquad g\rightarrow \infty.
\eea
The mass formula \eqref{mass}  follows by evaluating the matrix element ${}^g\langle 0|\rho|0\rangle^{g}$ in these two different representations for $\rho$, and taking into account that in the Abelian case $\langle \C|\C\rangle=\D^g$,
\bea
\label{mass1}
\NC(\T)= \sum_\C 1=\lim_{g\rightarrow \infty} {\D^g\over |\Mg|} \sum_{\gamma \in \Mg} {}^g\langle 0|U_\gamma|0\rangle^g.
\eea
In the end it simply evaluates the dimension of the modular-invariant subspace of $\H^g$ when $g\rightarrow \infty$. 

The sum on the RHS of \eqref{mass1}  is  the genus reduction of the Poincare series of the vacuum state $|0\rangle^g$ for an infinitely large genus $g$. It also can be interpreted as the TQFT partition function summed over all possible genus-$g$ Heegaard splittings of closed 3d manifolds. Indeed, the vacuum state $|0\rangle^g$ is the TQFT path integral on a given handlebody $X_3$, while  ${}^g\langle 0|U_\gamma|0\rangle^g \D^{g-1}$ evaluates the TQFT partition function on a connected closed 3d manifold obtained by gluing $X_3$ to itself after a mapping class group transformation $\gamma$.   
 It is well-known that all closed 3d manifolds can be obtained this way in the limit $g\to\infty$ \cite{hempel20223}.  
 We interpret the
RHS of \eqref{mass1} as the renormalized partition function of TQFT gravity -- a model of topological gravity based on a given TQFT summed over all 3d topologies.  

There is also a 
version of \eqref{mass1} that involves a sum over all 3d geometries with fixed boundary $\Sigma_g$, that expresses the holographic duality of TQFT gravity with an ensemble of boundary theories living on $\Sigma_g$ \cite{Dymarsky:2024frx,Aharony:2023zit,Dymarsky:2025agh}. 

An alternative way to calculate the number of TBCs is given by 
\bea
\label{trH}
\NC(\T)=\lim_{g\rightarrow \infty} {1\over |\Mg|}\sum_{\gamma \in \Mg}  {\rm Tr}_{\H^g}\left(U_\gamma\right). 
\eea
The RHS can be interpreted as the trace of the ``Wheeler–DeWitt'' time evolution operator in TQFT gravity. Since the latter is equal to the identity, its trace is equal to the dimension of the Hilbert space of a closed ``baby universe.''
Indeed the TBCs $\C$ should be understood as defining $\alpha$-states in the sense of \cite{Marolf:2020xie}; see \cite{Barbar:2025vvf} for a detailed discussion. 

The discussion above applies to Abelian theories. The extension of these observations to non-Abelian theories is postponed until section \ref{nonAbelian}.  

\subsection{Modular transformations and codes}
\label{modularcodes}
To evaluate the sum over topologies in \eqref{mass1}  we need to better understand the action of the mapping class group on the vacuum state. 
In theories admitting topological boundary conditions, the action of $U_\gamma$ on  $\H^g$ simplifies. It is convenient to introduce a new basis defined in terms of a given Lagrangian subgroup (even self-dual code) $\C_0\subset \mathscr{D}$ as follows
\bea
\label{newbasis}
\kket{(a,b)}:={1\over \D^{g/2}}\sum_{\vec{c}\, \in {\cal C}_0^g}
e^{i\pi\sum_i \scalar{a_i,b_i}+2\pi i \sum_i \scalar{a_i, c_i}}|\vec{c}+\vec{b}\rangle,\quad 
\eea
where $a_i,b_i,c_i\in \mathscr{D}$ and $i$ runs from $1$ to $g$, so that $(\vec a,\vec b)\in \mathscr{D}^{2g}$.
If the order of $\mathscr{D}$  is even, the phase factor $e^{i\pi \sum_i \scalar{a_i,b_i}}$ may not be well-defined. In this case $a_i,b_i$ should be understood as elements of the double-cover $\tilde{\mathscr{D}}$ of $\mathscr{D}$. An explicit example will be considered below in section \ref{toriccode}. 

The basis  \eqref{newbasis} is overcomplete. Vectors that correspond to the same equivalence classes $[\vec{a}],[\vec{b}]\in \mathscr{G}^g$, where $\mathscr{G}=\mathscr{D}/{\cal C}_0$, are equal to each other up to a phase.  
In terms of the new basis the original vacuum state can be written as follows
\bea
\label{0}
|0\rangle^g={1\over \D^{g/2}}\sum_{[\vec{a}] \in \mathscr{G}^g} \kket{(a,0)}.
\eea
This expression is independent of the choice of representatives of $[\vec{a}]$ inside $\mathscr{D}^{g}$.

The advantage of the new basis is that it simplifies the action of the mapping class group, 
\bea
\label{modular}
U_\gamma\kket{(a,b)}=\kket{\gamma(a,b)},
\eea
where $\gamma(a,b)$ denotes the fundamental action of $\gamma\in \Sp(2g,\Z)$ on the vector $(a,b)$ of length $2g$. 
From here it readily follows that the mapping class group acts on $\H^g$ as $\Sp(2g,\Z_k)$ where $k$ is the 
exponent of $\mathscr{D}$ (twice the exponent if $a_i,b_i \in \tilde{\mathscr{D}}$), which is the Frobenius-Schur exponent of 
 $\mathcal T$, the smallest $k$ such that $(\theta_c)^k=1$ for all $c$.  The expression  \eqref{0} also makes  manifest that $\Gamma_0(k)\subset \Sp(2g,\Z)$ is the stabilizer of the vacuum state. 
 The opposite is also true: if $\Gamma_0(k)\subset \Sp(2g,\Z)$ is the stabilizer of the vacuum state then the Abelian theory $\T$ admits TBCs, as we show in Appendix \ref{proof}.

We end this section with a side comment on the interpretation of these structures in terms of codes.
When $\mathscr{D}$ is the direct sum of $ \mathscr{G}$ and $\C_0$,  the construction further simplifies. 
The decomposition  $\mathscr{D}= \mathscr{G}\oplus \C_0$ is not universal, and  may depend on the choice of $\C_0$.
Any theory  of the form $\T \times \bar \T$ admits such a decomposition  when $\C_0$ is the diagonal invariant.  
Provided  $\mathscr{D}= \mathscr{G}\oplus \C_0$, 
 one can restrict $a_i,b_i$ to be elements of $ \mathscr{G}$; then $k$  is  the  exponent of $ \mathscr{G}$ (or  in certain cases twice larger). In this case the vacuum state \eqref{0} and its images under $\Sp(2g,\Z)$,  up to normalization,  are code states associated with classical symplectic  codes $\S\subset \mathscr{G}^{2g}$ of length $2g$,\footnote{In \cite{Dymarsky:2024frx, Dymarsky:2025agh,Barbar:2025krh} such codes were denoted by $\cal L$. We use $\S$ to specify these are classic {\it symplectic} codes.} 
\bea
\label{symplcodes}
&&\sum_{i=1}^g \scalar{a_i,b_i'}-\scalar{a_i',b_i}=0\, \, {\rm mod}\, \, 1,\qquad 
(a,b),(a',b')\in \S,\\
&&|\S\rangle=\sum_{(a,b)\in \S}\kket{(a,b)}. \label{State} 
\eea
We note, that both \eqref{Cstate} and \eqref{State} are quantum stabilizer states of the  Calderbank–Shor–Steane (CSS) type defined in terms of classical codes $\C$ and $\S$ correspondingly.  
In the general case $\S$ should be understood as a symplectic self-dual code within $\mathscr{D}^{2g}$ that defines the stabilizer state $|\S\rangle$.  The language of codes and stabilizer states is especially helpful when discussing the path integral on a 3-manifold with boundary, see \cite{Dymarsky:2024frx,Aharony:2023zit,Dymarsky:2025agh}.

\section{Sum over closed 3d manifolds}
\label{sum3d}
Our goal is to evaluate the sum  over topologies in \eqref{mass1}. For a theory $\T$ with  Frobenius-Schur  exponent $k$, it can be rewritten as a limit of finite sums
\bea
\NC(\T)&=&\lim_{g\rightarrow \infty}  \M_g,
\eea
where the ``genus-$g$ mass'' is
\bea
\M_g&=& {\D^g\over |\Sp(2g,\Z_k)|} \sum_{\gamma\in \Sp(2g,\Z_k)} {}^g\langle 0|U_\gamma|0\rangle^g.
\label{mainformula}
\eea
It is a well-known result that any Abelian theory $\mathcal T$ factorizes into a product of theories ${\mathcal T}_{p_i}$ labeled by distinct primes $p_i$ \cite{Cano:2013ooa}. 
As follows from \eqref{modular} and the Chinese Reminder Theorem, the 
action of the mapping class group  factorizes accordingly,   
\bea
\Sp(2g,\Z_k)=\prod_i \Sp(2g,\Z_{k_i}),\qquad k=\prod_i k_i,
\eea
with $\gamma \in \Sp(2g,\Z)$ being mapped to $\gamma\,{\rm mod}\, k_i$,  and  $k_i$ denote the Frobenius-Schur exponents of ${\mathcal T}_{p_i}$.
From here it immediately follows that the sum over  $\Sp(2g,\Z_k)$ in \eqref{mass1} factorizes, leading to the well-known result \cite{kaidi2022higher}
\bea
\label{product}
\NC(\T)=\prod_i \NC(\T_{p_i}).
\eea
In what follows we therefore can  focus on an individual theory $\T_p$  with Frobenius-Schur exponent $k=p^m$,  for $p$ prime.

We also note that when a theory is a product of $n$ equivalent copies $\T\rightarrow \T^{\otimes n}$,  the mass formula  adjusts in a simple way, with $\D\rightarrow \D^n$ and 
\bea
{}^g\langle 0|U_\gamma|0\rangle^g \rightarrow \left({}^g\langle 0|U_\gamma|0\rangle^g\right)^n.
\eea
In other words,  the results for any particular theory can be easily extended to the tensor product of $n$ copies. 

In any theory admitting topological boundary conditions and with Frobenius-Schur  exponent $k$,  the  vacuum is stabilized by $\Gamma_0(k)$. Hence the average in \eqref{mainformula} may be taken over the double-coset 
\bea
\label{doublecoset}
\coset_g=\Gamma_0 \backslash \Sp(2g,\Z_k) / \Gamma_0.
\eea
To illustrate the power of this decomposition  
we calculate the partition function of a theory $\T_p$ with $k=p^m$ on the lens space $\bar{L}(n,1)$:
\bea
\label{lensspace}
Z_{\mathcal T}[\bar{L}(n,1)]=\langle 0|U_{\gamma_n}|0\rangle=\langle 0|S^\dagger T^n S|0\rangle=\D^{-2}{\sum_{c\in \mathscr{D}} (\theta_c)^n }, \quad \gamma_n=\left(\begin{array}{cc}
1 & 0\\
n & 1\end{array}\right).
\eea
The phase of $Z_{\mathcal T}[\bar{L}(n,1)]$ defines the higher central charge  \cite{Ng_2019,Ng_2022,kaidi2022higher}
\bea
\zeta_n={\sum_c (\theta_c)^n\over \left|\sum_c (\theta_c)^n\right|}.
\eea
The conventional chiral central charge corresponds to $n=1$. 
When $n=0\,\,{\rm mod}\, \, k$, $\gamma_n$ is in the same equivalence class in $\coset_1$ as the identity element ${\bf 1}$, and when $n$ and $p$ are co-prime,  it is in the same equivalence class as ${\bf s}$,  where
\bea
{\bf  1}=\left(\begin{array}{cc}
1 & 0\\
0 & 1\end{array}\right),\, {\bf s}=\left(\begin{array}{cc}
0 & -1\\
1 & 0\end{array}\right) \in \SL(2,\Z_k).
\eea
This implies
\bea
Z_{\mathcal T}[\bar{L}(n,1)]=Z_{\mathcal T}[\mathbb{S}^2\times\mathbb{S}^1]=1,\quad k \mid n,
\eea 
 and 
 \bea
 Z_{\mathcal T}[\bar{L}(n,1)]=Z_{\mathcal T}[\mathbb{S}^3]=\D^{-1},\quad p \nmid n.
 \eea 
From here we immediately find that when $k$ divides $n$ or $p$ does not divide $n$, the higher central charges, including the chiral central charge, must vanish. By decomposing a general theory into a product $\T=\prod_i \T_{p_i}$ and applying this result to each $\T_{p_i}$ we rederive the following result of \cite{kaidi2022higher}: in an Abelian bosonic theory admitting topological boundary conditions, if $n=0$  mod $k_i$ or $p_i\nmid n$ 
for each $p_i$, where $k_i=p_i^{m_i}$, all central charges must vanish. As explained in \cite{kaidi2022higher} the same condition can be equivalently written as 
\bea
\label{ncondition}
{\rm gcd}\left(n,{2|\mathscr{D}|\over {\rm gcd}(n,2|\mathscr{D}|)}\right)=1.
\eea
The converse is also true -- trivial central charges for all $n$ satisfying \eqref{ncondition} imply $\NC(\T)>0$. We prove this in Appendix \ref{proof}.

Going back to $\coset_g$ for arbitrary $g$ and $k=p^m$ for prime $p$,  the coset representatives  can be chosen in the form
\bea
\gamma_h(X)= {\bf e}(X) \oplus {\bf s}^{g-h},\quad 
\label{1X}
{\bf e}(X)=\left(
\begin{array}{cc}
 \1_h & 0 \\
 p\, X & \1_h \\
\end{array}
\right), 
\eea
where $X$ is an arbitrary symmetric $h \times h$ matrix with values in $\Z_{p^{m-1}}$.
In the expression above we  assume the canonical embedding $\Sp(2h,\Z_k)\times \Sp(2(g-h),\Z_k) \subset \Sp(2g,\Z_k)$.
When $k=p$ is prime,  the representatives simplify to  $\gamma_h ={\bf 1}^{h} \oplus {\bf s}^{g-h}$.  Hence, for $k=p$ the partition function of $\T_p$ on any connected closed 3-manifold is equal to $\T_p$ on the connected sum of $h$ copies of $\mathbb{S}^2\times\mathbb{S}^1$ or, when $h=0$, on $\mathbb{S}^3$. 

The representatives $\gamma_h(X)$ are not uniquely defined. Congruent choices of $X$,
\bea
X \sim A\, X\, A^T, \qquad A\in \GL(h,\Z_k),
\eea
belong to the same orbit and yield the same matrix element
\bea
f(X)\equiv {}^h\langle0|U_{\gamma(X)} |0\rangle^{h},\qquad \gamma(X)={\bf e}(X).
\eea

We define $\Sa$ to be the subgroup of $\Gamma_0$ that preserves the form of $\gamma_h$, while possibly changing $X$.
Its size is evaluated in Appendix \ref{groups},
\bea
&&|\Sa|=|\GL(h,\Z_k)| |\GL(g-h,\Z_k)| k^{2h(g-h)+h(h+1)/2}. \label{Sa} 
\eea
We now introduce   the measure 
\bea
&&\mu(h,g,p)={|\Gamma_0(2g,\Z_k)|^2 p^{(m-1)h(h+1)/2} \over |\Sp(2g,\Z_k)| |\Sa|}=
{p^{(g-h)(g-h+1)/2} \over \prod_{i=1}^g (p^i+1)}
\prod_{i=1}^h \frac{p^{g-i+1}-1}{p^i-1}, \label{measure}
\eea
which is $m$-independent, see Appendix \ref{measureApp} where this point is further explained. 
In terms of $\mu$ the sum over topologies is 
\bea\label{Mg}
 \M_g  =\sum_{h=0}^g \mu(h,g,p)\, \D^h\, \G_h,\quad \G_h=p^{(1-m)h(h+1)/2}  \sum _X\,  f(X).
\eea 
There are several different ways to think about $f(X)$ and  the sum  over $X$ in $\G_h$. One is  discussed in Appendix \ref{measureApp}.
In another interpretation, $\G_h$ is a generalization of the lens space partition function discussed above
\bea
\label{fXdef}
f(X)= {}^h\langle0|U_{\gamma(X)} |0\rangle^{h}={1\over \D^{2h}} \sum_{c_i\, \in \mathscr{D}} e^{i \pi \sum_{ij} p\scalar{c_i,c_j} X_{ij}}.
\eea
We can think of $p\scalar{c_i,c_j} X_{ij}$ as defining a quadratic form on $\mathscr{D}^g$, the classification of which is known. Hence  $f(X)$ is a combination of standard Gauss sums. The nontrivial task is to express it in terms of invariants of $X$ to allow for explicit evaluation of $ \G_h$ for an arbitrary product of theories $\T_p \times \T_{p}' \times \dots$.  For odd $p$ we will carry out this procedure below in a series of examples. 

Computationally, to evaluate $ \G_h$ it is usually easier  to first perform the sum over $X$. When $p$ is odd this will  enforce the conditon 
that a combination of  $\{c_1,\dots, c_g\}$ contributes only if 
\bea
p \scalar{c,c}=0\,\, {\rm mod}\, \, 2, \label{evennessp}
\eea
where $c$ is an arbitrary linear combination of the $c_i$. In other words $c_i$ are the codewords of an even additive code within $\mathscr{D}$, where evenness is defined by \eqref{evennessp}, c.f.~\eqref{evenness}. This way of thinking is helpful in simple cases, and it immediately shows that $ \G_h$ is non-negative and real,   but counting codewords becomes cumbersome when $\T$ is a product of many different theories.

Once $\G_h$ has been evaluated, we can take the limit $g\rightarrow \infty$ and express the formula for 
mass  in the following form 
\bea
&&\NC(\T_p)=\sum_{h=0}^\infty \mu(h,p)\, \D^h\, \G_h, \quad \G_h=p^{(1-m)h(h+1)/2} \sum _X\,  f(X),\\
&&\mu(h,p)=\lim_{g\rightarrow \infty} \mu(h,g,k)={p^{-h(h+1)/2}\over 
 \prod_{i=1}^\infty (1+p^{-i}) \prod_{i=1}^h (1-p^{-i})}. \label{muinf}
\eea
Together with \eqref{product} this provides a universal way to evaluate the number of TBCs for any Abelian bosonic theory.
We now proceed evaluating  $\G_h$ in a series of examples.  


\subsection{Level-$k$ toric code}
\label{toriccode}
Our first example is the $n$ copies of the ``level-$k$ toric code''  Chern-Simons theory with K-matrix $k\, \sigma_x$, otherwise known as $\Z_k$ Dijkgraaf-Witten theory.  The discriminant group of $n$ copies of this  theory is $\mathscr{D}=(\Z_k \times \Z_k)^n$, with the scalar product 
\bea
\label{qtoric}
\scalar{c,c}=2q(c),\quad q(c)=\sum_i {\alpha_i \beta_i\over k},\qquad c=(\alpha_1,\dots,\alpha_n,\beta_1,\dots,\beta_n)\in \Z_k^{2n}.
\eea
This theory has vanishing chiral central charge and admits TBCs for any $n$. The Lagrangian subgroups $\C$ are the subgroups of $\Z_k^{2n}$ of order $k^n$ satisfying  \eqref{evenness}.
For $p=2$ these conditions define a subclass of codes introduced in the pioneering work of  \cite{Calderbank:1996aj} on quantum stabilizer codes.
In the classification of \cite{Nebe} these are codes of type $4^{H+}$ -- self-dual codes over $F_4$ with Hermitian inner product.  The condition of evenness additionally imposes that, understood as quantum stabilizer codes over qubits, these codes are real. These codes were studied extensively in the context of code CFTs in \cite{Dymarsky:2020bps,Dymarsky:2020qom}. The generalization to arbitrary $p$ was introduced in \cite{Yahagi:2022idq,Angelinos:2022umf} and discussed in the context of the holographic duality between Abelian TQFT gravity and ensembles of Narain CFTs in \cite{Aharony:2023zit,Dymarsky:2025agh}.
 
The level $k$  is also the Frobenius-Schur exponent of this TQFT.  For any $n$ this theory admits TBCs, including the ``Dirichlet'' one $\C_0=(\alpha_1,\dots,\alpha_n,0,\dots 0)$ where $\alpha_i\in \Z_k$. The quotient $ \mathscr{G}=\mathscr{D}/\C_0$ satisfies $\mathscr{D}= \mathscr{G}\oplus \C_0$ and $a_i,b_i$ in \eqref{newbasis} are elements of $\mathscr{G}=\Z_k^{n}$. This basis  is non-degenerate; it is related to the conventional basis $|(\alpha,\beta)\rangle$ \eqref{basis} by a Fourier transform with respect to $a$ \cite{Barbar:2025krh}, and it linearizes the action of $\Sp(2g,\Z_k)$. 


We proceed by evaluating the number of codes $\C$, starting 
 with the case of prime $k=p$,
for which the mass for genus $g$ is given by 
\bea\label{toricmass}
 \M_g
 =\sum_{h=0}^g \mu(h,g,p)\, \D^h=\prod_{i=1}^g  {p^i+p^n\over p^i+1}=p^{gn}\prod_{i=0}^{n-1} {p^i+1\over  p^{i}+p^g},\quad \D=p^n.
\eea 
The last equality is proved in Appendix \ref{identityproof}.
After taking the $g\rightarrow \infty$ limit the measure simplifies, yielding the mass formula for the theory $\T_p$ with prime Frobenius-Schur exponent,
 \bea
 \NC(\T_p)&=&\sum_{h=0}^\infty \mu(h,p)\, p^{hn}=\sum_{h=0}^\infty {p^{hn -h(h+1)/2}\over 
 \prod_{i=1}^\infty (1+p^{-i}) \prod_{i=1}^h (1-p^{-i})}\nonumber\\ 
&=& \prod_{i=0}^{n-1}(p^i+1)=|\Or(n,n,\Z_p)/\Gamma_0(n,n,\Z_p)|.
 \label{toriccodeN}
\eea
The coset expression 
for $ \NC(\T_p)$ was first given in \cite{Gaiotto:2020iye} in the context of orbifold groupoids and in \cite{Dymarsky:2020qom,Aharony:2023zit} in the context of code CFTs. 
We briefly review the coset construction in Appendix \ref{orthogonal}. The resulting expression is well-known in the coding literature as giving the number of self-dual codes over $\Z_p$ \cite{pless1965number}. 

Before we proceed with the case of composite $k$, 
we note that the theories consisting of $n$ copies of the level-$p$ toric code are the only TQFTs with odd-prime Frobenius-Schur exponent. This will be proven below.  

When $k=p^2$, we need to evaluate 
\bea
f(X)={1\over p^{4h}}\sum_{\alpha,\beta \in \Z_k^{^h}} e^{{2\pi i \over p} \alpha^T X \beta}.\label{fx}
\eea
We can write $\alpha=\alpha_0 +p \alpha_1$ and similarly for $\beta$, where $\alpha_0,\alpha_1\in \Z_p^h$. The expression in \eqref{fx} is independent of $\alpha_1,\beta_1$; hence the corresponding sum is trivial. We are left with 
\bea
f(X)={1\over p^{2h}}\sum_{\alpha_0,\beta_0 \in \Z_p^{^h}} e^{{2\pi i \over p} \alpha_0^T X \beta_0}= p^{-r},
\eea
where $r={\rm rank}(X)$.  
The number of $h\times h$ symmetric matrices of rank $r$ with values in $\Z_p$ is given by \cite{rankR}
\bea
 \label{Nar}
N(h,r,p)&=&{\prod_{i=1}^h (1-p^{-i})\over \prod_{i=1}^r (1-p^{-i}) \prod_{i=1}^{h-r} (1-p^{-i})} p^{h r-r(r-1)/2}\prod_{i=0}^{[(r-1)/2]
}\left(1-p^{-(2i+1)}\right),\nonumber
\eea
This formula applies to any prime $p$.  For odd $p$ it follows from $N(h,r,p)=\sum_{\ell=0}^1 N(h,r,\ell, p)$, where
\bea
\nonumber
N(h,r,\ell, p)&=& {|\GL(h,\Z_p)|\over |\Or_{\ell}(r,\Z_p)| |\GL(h-r,\Z_p)|\, p^{r(h-r)}}\\ 
&=&N(h,r,p){1+(-1)^{\ell+(p-1)/2}\, p^{-r/2} \delta_{2 | r } \over 2},  \label{Narl}
\eea
is the number of quadratic forms of rank $r$ whose nondegenerate part has determinant $\epsilon^\ell$, $\epsilon$ being a non-square element in $\Z_p$, and $\Or_\ell$ is the symmetry group of the corresponding quadratic form.

We therefore find   the number of TBCs for $n$ copies of the toric code of level $k=p^2$ to be 
\bea
 \nonumber
 \NC(\T_{k=p^2})&=&\sum_{h=0}^\infty \mu(h,p)\, p^{2hn-h(h+1)/2} \sum_{r=0}^h  N(h,r,p)\, p^{-nr}\\ \label{toricp2}
 &=&\sum_{\tilde{n}=0}^n p^{\tilde{n}(\tilde{n}-1)/2} \sigma(\tilde{n}),\quad \sigma(\tilde{n})=\prod_{i=1}^{\tilde{n}} {(p^{n-i}+1)(p^{n+1-i}-1)\over (p^i-1)},
\eea
in agreement with \cite{Barbar:2025krh}. Here $\sigma(\tilde{n})$ is the number of even codes $[n,\tilde{n}]$ of length $n$ and with $\tilde n$ generators. 
The same expression was previously found in the coding literature \cite{BALMACEDA20082984}. 

We can now consider the case of $k=p^m$ for arbitrary $m$.  The matrix $X$ in this case can be decomposed as follows
\bea
X=X_0+p X_1+\dots p^{m-1} X_{m-1},
\eea
and similarly $\alpha=\alpha_0+p\alpha_1+\dots $. and $\beta=\beta_0+\dots$. 
Using transformations $X \rightarrow A X B$ with $A,B\in \GL(h,\Z_k)$ that do not change $f(X)$, we can bring $X_i$ to the following form. We first diagonalize $X_0$, such that its top left block is an invertible  diagonal matrix $D_{r_0}$ of size $r_0={\rm rank}(X_0)$.  We then leave the first $r_0$ rows and columns of $X_1$ arbitrary, while diagonalizing the rest, as shown in \eqref{Xmat}.  There,  $r_1$ denotes  the rank of the bottom right submatrix of $X_1$ of size $h-r_0$. We then leave first $r_0+r_1$ columns and rows of $X_2$ arbitrary and diagonalize the rest, etc. 
\bea
\label{Xmat}
X_0 = 
\begin{pmatrix}
D_{r_0} & 0 & 0  & 0\\
0 & 0 &0  & 0\\
0 & 0 &0  & 0\\ 
0 & 0 &0  & 0
\end{pmatrix},
\quad
X_1 =
\begin{pmatrix}
* & *  &  * & *\\
* & D_{r_1} & 0 &0\\
* & 0 & 0 & 0 \\
* & 0 & 0 & 0 
\end{pmatrix},\quad
X_2 =
\begin{pmatrix}
* & *  & * & *\\
* & *  & * & *\\
* & * & D_{r_2} & 0\\
* & * & 0  & 0
\end{pmatrix},\dots 
\eea
To evaluate the sum $f(X)=\sum_{\alpha,\beta} e^{2\pi i \alpha^T X \beta/p^{m-1}}/p^{2hm}$, we  note that 
the exponent is not dependent on $\alpha_{m-1},\beta_{m-1}$. We then sum over $\alpha_{m-2},\beta_{m-2}$ which will force the first $r_0$ entries of $\alpha_0,\beta_0$ to be zero. Summing over the components $r_0+1,\dots, r_0+r_1$ of
$\alpha_{m-3},\beta_{m-3}$ will force next $r_1$ components of $\alpha_0,\beta_0$ to be zero, etc. In the end, the last $(h-r_0-r_1-\dots -r_{m-3})$ components of   $\alpha_0,\beta_0$ will enter the sum $e^{2\pi i \alpha_0^T X_{m-2} \beta_0/p}$, yielding $p^{-r_{m-2}}$.
Combining everything together we find
\bea
f(X)= p^{-\sum_{i=0}^{m-2}(m-1-i)r_i}, 
\eea
and the number of TBCs for $n$ copies of the level $k=p^m$ toric code is given by 
\bea
\label{tcN}
&& \NC(\T_{k=p^m})=\sum_{h=0}^\infty \mu(h,p)\, p^{h m n}\, \G_h,\\ \nonumber 
&& \G_h= \sum_{r_0=0}^h  p^{-h(h+1)/2}N(h,r_0,p) \sum_{r_1=0}^{h-r_0} p^{-(h-r_0)(h-r_0+1)/2}N(h-r_0,r_1,p)\times \\  \nonumber 
&& \dots \times    \sum_{r_{m-2}=0}^{h-\tilde{r}}  p^{-(h-\tilde{r})(h-\tilde{r}+1)/2} N(h-\tilde{r},r_{m-2},p) \, p^{-n \sum_{i=0}^{m-2}(m-1-i)r_i}, \quad \tilde{r}=r_0+\dots +r_{m-3}.
\eea
Similarly to the $k=p^2$ case \eqref{toricp2},  this expression should admit a representation as a sum over $\Z_{k}$ codes as discussed in \cite{nagata2008constructive,nagata2009number,nagata2013mass}.

\subsection{$\T_p$ with  odd prime $p$ and $k=p^m$}
We now turn to a general theory $\T_p$ for an odd prime $p$. To begin with, we assume that $k=p$, which means $\mathscr{D}=\Z_p^{\tilde n}$ for some $\tilde{n}$. We first assume $\tilde{n}=1$ and then consider a product of such theories. 
For any odd prime $p$,  there are two different quadratic forms \cite{Cano:2013ooa}
\bea
q(c)=u {c^2\over p},\quad c\in \Z_p,
\eea
where $u=1$ or $u=\epsilon$  is a non-square element  in $\Z_p$. When $p=3\, {\rm mod}\, 4$, we can choose $\epsilon=-1$. When $p=1\, {\rm mod}\, 4$, there is no canonical choice of $\epsilon$.  When  $p=3\, {\rm mod}\, 4$, theories with $u=\pm 1$ have chiral central charge $c_-=\pm 2$. 
When $p=1\, {\rm mod}\, 4$, the theory with $u=1$ has chiral central charge $c_-=0$ and the theory with $u=\epsilon$ has chiral central charge $c_-=4$. 

Now we consider the case of arbitrary $\mathscr{D}=\Z_p^{\tilde n}$. For the theory to  admit a TBC, it has to be even $\tilde{n}=2n$, and $\D=p^n$. 
Without loss of generality we can  diagonalize the quadratic form as
\bea
\label{qfp}
q(c)={1\over p}\left(\sum_{i=1}^{2n-1} c_i^2 +u\, c_{2n}^2\right),\qquad \vec{c}\in \Z_p^{2n},
\eea
where $u$ is as above.
Since, by assumption, the theory admits TBCs, its chiral central charge should vanish mod $8$. For $p=1\,\, {\rm mod}\, \, 4$ the total central charge is 
\bea
c_- =\left\{ \begin{array}{l l}
2n\times 0, & {\rm for}\, \,  u\, \, {\rm square},\\
(2n-1)\times 0 +4,\, & {\rm for}\, \,  u\, \, {\rm nonsquare},
\end{array}\right.
\eea 
hence  $u$ is a square.
For $p=3\,\, {\rm mod}\, \, 4$ the total central charge is 
\bea
c_-=\left\{ \begin{array}{ll}
2n\times 2, & {\rm for}\, \,  u\, \, {\rm square},\\
(2n-1)\times 2 -2,\, &{\rm for}\, \,  u\, \, {\rm nonsquare}.
\end{array}\right.
\eea 
Hence $u$ is a square when $n$ is even and non-square when $n$ is odd. In all cases we find the determinant of the quadratic form \eqref{qfp}  to be $(-1)^n$, which means it can be brought to the form 
\bea
q(c)={1\over p} c^T \eta\,  c,\qquad \eta=\left(\begin{array}{cc}0 & 1_n \\
1_n & 0\end{array}\right).
\eea
In other words, the only theory $\T_p$  admitting TBCs with Frobenius-Schur exponent $k=p$, where $p$ is an odd prime, is $n$ copies of the level $k=p$ toric code. 

The Lagrangian subgroups of this theory $\C \in \Z_p^{2n}$ can be readily interpreted as self-dual codes of length $2n$ over $\Z_p$ with scalar product
\bea
\sum_{i=1}^{2n-1} c^{}_i c'_i +u\, c^{}_{2n} c'_{2n}=0\,\, {\rm mod}\, \, p,\qquad c,c'\in \Z_p^{2n}.
\eea
Since $2$ is invertible in $\Z_p$ ``evenness'' \eqref{evenness} follows from self-duality. The case of $u=1$ is most natural; such codes have been discussed extensively in the literature \cite{PLESS1975313,pless1965number}.
We note that the straightforward application of Construction A applied to a self-dual code $\C$ would yield a self-dual but odd lattice 
\bea
\Lambda_\C=\left\{ \vec{v}/\sqrt{p}\, |\, \vec{v}=(c_1,\dots,c_{2n}),\, c_i\in \Z,\, \, \vec{v}\,\, {\rm mod}\, \,  p\in \C\right\}\subset \R^{2n}.
\eea
To obtain an even self-dual lattice, that defines a Narain CFT in terms of the sandwich construction of bosonic CS theory \cite{Barbar:2023ncl}, one should use the modified Construction A outlined in \cite{Angelinos:2022umf}, which is defined in terms of the lattice $\Lambda$ of the underlying CS theory. 

We now turn to theories $\T_{k=p^2}$ with the Frobenius-Schur exponent $k=p^2$.  Similarly to the above, when $\mathscr{D}=\Z_{k}$ there are two quadratic forms 
\bea
q(c)=u {c^2\over p^2},\quad u=\epsilon^l,\quad  c\in\Z_{p^2}, \label{theoryp2}
\eea
The corresponding $K$ matrix is given by 
\bea
K=\left(\begin{array}{cc}
2u & p \\
p & 0
\end{array}\right).
\eea
For both possible values of $u$, this theory has vanishing chiral central charge. 
Theories with $\mathscr{D}=\Z_{k}^n$ can be obtained as a product of theories of these two types, with  $u=1$ for the first $n-1$ theories  and $u=1$ or $u=\epsilon$ for the $n$-th theory.  

We now want to evaluate 
\bea
\label{fXp2}
f(X)={1\over k^h} \sum_{c\in \Z_k^h} e^{2\pi i {u\over p} c^T X c}.
\eea
A symmetric matrix $X$ with values in $\Z_p$ can be diagonalized as
\bea
\label{Xdiag}
X(r,\ell)={\rm diag}(\underbrace{1,\dots,1,u_x}_{r},0,\dots,0), 
\eea
where $u_x$ is either $1$ or a non-square element $u_x=\epsilon$, which can be written 
as $u_x=\epsilon^{\ell}$ for $\ell=0,1$. The sum in \eqref{fXp2} is now a product of standard Gauss sums  yielding
\bea
f(X)=f(l,r,\ell)= (-1)^{rl+\ell}\, i^{(p-1)^2 r\over 4} p^{-r/2}. 
\eea
To evaluate $\G_h$  for $n$ copies of the theory \eqref{theoryp2}, we note that the total number of matrices that can be brought to the form \eqref{Xdiag} specified by $r,\ell$  is given by \eqref{Narl} and 
\bea
&&\G_h=\sum_{r=0}^h \sum_{\ell=0}^1 p^{-h(h+1)/2} N(h,r,\ell, p) f^n(r,\ell). \label{Gan}
\eea
Combining everything together we find that the number of TBCs is
\bea
\nonumber 
 \NC(\T^n_{k=p^2})&=&\sum_{h=0}^\infty \mu(h,p)\, p^{hn} \G_h.
\eea
When $p=1\,\, {\rm mod}\, \, 4$, the theory $\T_{k=p^2}$ and its conjugate $\overline \T_{k=p^2}$ are the same. Hence when $n$ is even, $\T_{k=p^2}^{\otimes n}$ is equivalent to $n/2$ copies of the level $k=p^2$ toric code.  This equivalence is manifest in  the fact that $f^2$ is independent of $\ell$. More generally, for any  $p$, the combination $\T_{k=p^2} \times \overline \T_{k=p^2}$ is equivalent to a single copy of the  level $k=p^2$ toric code. 

It is interesting to note that one copy of the $\T_{k=p^2}$ theory has exactly one TBC, hence  for any $g$ the mass is equal to one. This readily follows from 
\bea
\M_g=\sum_{h=0}^g \mu(h,g,p)=1, 
\eea
where we used the fact that  $\G_h=p^{-h}$, which follows from \eqref{Gan} with $n=1$, or which can be evaluated directly from \eqref{fXp2} by first averaging over $X$. 
This theory provides the simplest example for which  $\mathscr{D}=\Z_{p^2}$ is not a direct sum $\mathscr{G}\oplus \C$, as follows from taking $\C=\Z_p$. As a result 
$a$ and $b$ in \eqref{newbasis} should be defined to be elements of $\mathscr{D}=\Z_{p^2}$. This basis is overcomplete by a factor of $p^2$, yet it linearizes the action of the mapping class group $\Sp(2g,\Z_{p^2})$.

We are now ready to consider the case of general $k=p^m$. Starting from the theory defined by the quadratic form 
\bea
q(c)=u {c^2\over p^m},\quad  u=\epsilon^l,\quad  c\in\Z_{p^m}, \label{theorypm}
\eea
we wish to evaluate 
\bea
f(X)={1\over k^h}\sum_{c\in \Z_k^a} e^{2\pi i {u p\over k } c^T X c}.
\eea
There are two ways to proceed.  The first is to   decompose $X=X_0+p X_1+p^2 X_2+\dots$ and $c=c_0+p c_1+p^2 c_2+ \dots$, and bring $X_i$ to the form outlined in  \eqref{Xmat} with each diagonal matrix $D_i={\rm diag}({1,\dots,1,\epsilon^{\ell_i}})$.  Then the sum can be done explicitly,  starting from $c_i$ with $i=m-2$ and going to $i=0$. It is readily seen that  the values in the first $r_0$ rows and columns  of $X_1$, as well as in the first $r_0+r_1$ rows and columns of $X_2$, etc., do not affect the answer and simply yield a factor $p^{r_0(r_0+1)/2+r_0(h-r_0)+\dots}$. 
This leads to 
\bea
\nonumber
\G_h&=&p^{(1-m)h(h+1)/2} \sum_{r_0}^h \sum_{\ell_0=0}^1 N(h,r_0, \ell_0, p) p^{r_0(r_0+1)/2+(h-r_0)r_0}\sum_{r_1=0}^{h-r_0} \sum_{\ell_1} N(h-r_0,r_1,\ell_1,p)\times \\ 
&& \dots \times   p^{\tilde{r}(\tilde{r}+1)/2+(h-\tilde{r})\tilde{r}} \sum_{r_{m-2}=0}^{a-\tilde{r}}  \sum_{\ell_{m-2}} N(h-\tilde{r},r_{m-2},\ell_{m-2},p)
f(X) ,\quad \tilde{r}=r_0+\dots + r_{m-3},\label{GgeneralT0} \nonumber
\eea
with
\bea
f(X)=f(l,\vec{r},\vec{\ell})= \prod_{i=0}^{m-2}  (-1)^{(r_i l+\ell_i) (m-1-i)} \, i^{(p-1)^2 r_i \delta_{m-1-i} \over 4}\,  p^{-(m-1-i)r_i},  
\label{fXgeneralT}
\eea
where $\delta_i=1$ if $i$ is odd and $\delta_i=0$ if $i$ is even.

Another way to obtain the same result is to bring $X$ to the diagonal form
\bea
\label{XmatD}
X=\begin{pmatrix}
D_{r_0} & 0 & 0 & 0 & 0 \\
0 & p\,D_{r_1} & 0 & 0 & 0 \\[4pt]
0 & 0 & \ddots & \ddots & 0 \\[4pt]
0 & 0 & \ddots & p^{\,m-2} D_{r_{m-2}} & 0 \\
0 & 0 & 0 & 0 & 0
\end{pmatrix}
\eea
where $D_{r_i}={\rm diag}(\underbrace{1,\dots,1,\epsilon^{\ell_i}}_{r_i})$. The number of  matrices  congruent to \eqref{XmatD}, parametrized by $r_0,r_1,\dots,r_{m-1}$ and $\ell_0,\dots, \ell_{m-2}$, is given by the size of the stabilizer of $X$ inside $\GL(h,\Z_k)$, 
 \bea
  &&N(h,r_0,r_1,\dots, r_{m-2},\ell_0,\dots,\ell_{m-2})=\\ \nonumber
 &&{|\GL(n,\Z_{p^{m-1}})|
 p^{-(m-1)r_0(n-r_0)- m r_1 (n-r_0-r_1) -\dots -(2m-3)r_{m-2}(n-r_0-\dots)}
 \over |\Or_{\ell_0}(r_0,\Z_{p^{m-1}})| |\Or_{\ell_1}(r_1,\Z_{p^{m-2}})|p^{r_1^2} |\Or_{\ell_2}(r_2,\Z_{p^{m-3}})|p^{2r_2^2}\dots |\GL(n-r_0-\dots,\Z_{p^{m-1}})| }.
 \eea
Note that $|\Or_{\ell_j}(r_j,\Z_{p^{m-1-j}})|p^{j r_j^2}$ is the size of the group of  $r_j \times r_j$ matrices $A$ with values in $Z_{p^{m-1}}$ such that 
$A (p^j D_j) A^T=p^j D_j\, \, {\rm mod}\,\, p^{m-1}$. After a straightforward simplification this can be written as 
\bea
\nonumber
\G_h&=&\sum_{r_0}^h \sum_{\ell_0=0}^1  p^{-h(h+1)/2}  N(h,r_0, \ell_0, p) \sum_{r_1=0}^{h-r_0} \sum_{\ell_1=0}^1  p^{-(h-r_0)(h-r_0+1)/2}N(h-r_0,r_1,\ell_1,p) \times \\
&& \dots \times  \sum_{r_{m-2}=0}^{h-\tilde{r}}  \sum_{\ell_{m-2}=0}^1  p^{-(h-\tilde{r})(h-\tilde{r}+1)/2+(h-\tilde{r})\tilde{r}} N(h-\tilde{r},r_{m-2},\ell_{m-2},p)
f(X). \label{GgeneralT}
\eea

This completes the analysis of a general theory $\T_p$ for an odd prime $p$.  The most general theory can be obtained as a product of theories with different Frobenius-Schur  exponents $k=p^{m_I}$ and  quadratic forms given by \eqref{theorypm}. The corresponding $\G_h$ for the resulting theory will be given by \eqref{GgeneralT} with $m=\max_I m_I$ and $f(X)$ being a product of \eqref{fXgeneralT} for each individual theory (this function will depend only on the first $m-2$ variables $r_i$ and $\ell_i$),
\bea
f(X)=\prod_I f(l_I, r_1,\ell_1,\dots, r_{m_i-2},\ell_{m_i-2}).
\eea

We complete the discussion of $\T_p$ theories for odd primes $p$ by noting that since a symmetric matrix $X$ with values in $\Z_{p^{m-1}}$ can always be diagonalized, 
the partition function of $\T_p$ on any connected closed 3-manifold is equal to that one on a connected sum  of lens spaces ${\bar  L}(n,1)$ for $n< k=p^m$ or on $\mathbb{S}^3$. 

\subsection{$k=2$ theories}
We now turn to the case $p=2$. The  only  theories  with $k=2$ are the level-$2$ toric code, the level-$1$ WZW ${\rm Spin}(8)$ theory, and their combinations. The  toric code was discussed above.  In terms of Abelian Chern-Simons theory,   ${\rm Spin}(8)_1$ can be defined by taking $K$ to be the Cartan matrix of $\SO(8)$.
This theory has chiral central  charge $c_-=4$. 
The corresponding discriminant group is $\mathscr{D}=\Z_2 \times \Z_2$ and 
\bea
\label{SO8}
q(c)=(\alpha^2+\alpha\beta+\beta^2)/2 ,\quad c=(\alpha,\beta)\in \mathscr{D}.
\eea
For any theory with $k=2$ the number of TBCs will be given by \eqref{toriccodeN} with $n=n_1+n_2$ where $n_1$ is the number of level $2$ toric codes and $n_2$ is the number of ${\rm Spin}(8)_1$. Note that $n_2$ has to be even; otherwise the resulting theory will have a chiral central charge not divisible by $8$.  

\subsection{$\Un(1)_2$}
\label{dec}
We next consider the $\Un(1)_2$ theory and its conjugate $\Un(1)_{- 2}$, two examples of theories with $k=4$.  We first  focus on the case of $n$ copies of $\Un(1)_2$. The corresponding quadratic form is 
\bea
q(c)={1\over 4}\sum_i c_i^2,\qquad c\in \Z_2^n.
\eea
The Lagrangian subgroups of $\Un(1)_2^n$ are subgroups  $\C\in \Z_2^n$ of size $2^{n/2}$  satisfying 
\bea
\sum_i c_i =0\, \, {\rm mod}\, \, 4,\qquad c\in \C.
\eea
This condition defines doubly-even self-dual binary codes, the so-called binary codes of type II,  which have been extensively discussed in the coding literature \cite{nebe2009self,conway2013sphere}.   It is well-known that such codes exist only when $n=0\, \, {\rm mod}\, \, 8$, which is also the condition for  $\Un(1)_2^n$ to admit a TBC.

Any doubly-even self-dual binary code defines, via Construction A,  an even self-dual lattice in $\R^n$. This construction was used in \cite{dolan1996conformal,DOLAN1990165}  to construct chiral lattice-based 2d CFTs. The interpretation of such a CFT as arising from a ``sandwich'' construction, with the code defining the topological boundary condition  in 3d Chern-Simons theory, was given in \cite{Barbar:2023ncl}.

The number of doubly-even self-dual binary codes, given by \eqref{masstypeII}, is well-known. We  reproduce this result below. 
We first evaluate
\bea
\label{fU2}
f(X)={1\over 2^h}\sum_{c\in \Z_2^h} e^{i\pi c^T X c}
\eea
which vanishes unless all diagonal elements of $X$ are zero. Thus, 
\bea
\label{Gval2}
\G_h=2^{-h(h+1)/2}\sum_X \prod_{i=1}^h \delta_{X_{ii},0}=2^{-h},
\eea
and (this identity is proved in Appendix \ref{identityproof})
\bea 
\label{numberofbinarycodes}
\NC(\Un(1)_2^n)=\sum_{a=0}^\infty {2^{-h(3+h)/2+nh/2}\over \prod_{i=1}^\infty (1+2^{-i}) \prod_{i=1}^h (1-2^{-i})} = \prod_{i=0}^{n-2} (2^i+1).
\eea

To be systematic and allow for the evaluation of $\NC$ for any theory $\T_2$ with $k=4$, we express $f(X)=\prod_{i=1}^h \delta_{X_{ii},0}$  in terms of invariants of $X$. When its rank $r$ is odd, $X$ can be brought to  diagonal form with an $r\times r$ identity block and zeros elsewhere. There are 
\bea
N(h,r,2)={|\GL(h,\Z_2)|\over |\Or(r,\Z_2)| |\GL(h-r,\Z_2)| 2^{r(h-r)}}
\eea
such matrices (this is the same expression as in \eqref{Nar} with $p=2$),
where $\Or(r,\Z_2)$ is defined as the group of all matrices $A\in \Or(r,\Z_2)$ preserving the bilinear form, $A A^T=1$,
\bea
\quad |\Or(r,\Z_2)|=2^{\tilde{r}^2}\prod_{i=1}^{\tilde{r}} (p^{2i}-1),\quad r=2\tilde{r}+1.
\eea

When $r$ is even, there are two cases to consider -- either $X$ diagonalizes as an $r\times r$ identity block (with zeros elsewhere) or as an $r\times r$ symplectic form $J$, represented as $r/2$ copies of $\sigma_x$. 
In the former case, which in our notation corresponds to $\ell=0$, the group $\Or_{0}$ is defined as  the group of matrices satisfying $A A^T=1$. In the latter case, corresponding to $\ell=1$, $\Or_{1}$ is the group of all symplectic matrices,
\bea
|\Or_{0}(r,\Z_2)|&=& 2^{{\tilde r}^2}\prod_{i=1}^{{\tilde r}-1} (p^{2i}-1),\\ 
|\Or_{1}(r,\Z_2)|&=&|\Sp(2{\tilde r},\Z_2)|=2^{{\tilde r}^2}\prod_{i=1}^{\tilde r} (p^{2i}-1).
\eea
The number of such matrices $X$ is then
\bea
N(h,r,\ell,2)={|\GL(h,\Z_2)|\over |\Or_\ell (r,\Z_2)| |\GL(h-r,\Z_2)| 2^{r(h-r)}},\quad r=2\tilde{r}.
\eea
If for odd $r$ we define $N(h,r,\ell,2)=(1-\ell)N(h,r,2)$, then for any $r$, 
\bea
N(h,r,2)=\sum_{\ell=0}^1 N(h,r,\ell,2).
\eea

The condition that all $X_{ii}$ vanish implies that for $\ell=1$,
\bea
\sum_{r=0}^{h} N(h,r,1,2) = \sum_{\tilde{r}=0}^{[h/2]} N(h,2\tilde{r},1,2)=2^{h(h-1)/2},
\eea
in agreement with \eqref{Gval2}.

For the $\Un(1)_2^{n} \times \Un(1)_{-2}^{\bar n}$ theory, the calculation will be exactly the same with $n\rightarrow n+\bar n$. The corresponding binary codes, which  generalize the conventional doubly-even codes by modifying the Hamming weight to
\bea
\label{modifiedweight}
w(c)=\sum_{i=1}^{n} c_i -\sum_{i=n+1}^{n+\bar n} c_{i},\quad c=(c_1,\dots,c_{n}|c_{n+1},\dots,c_{n+\bar n}),
\eea
were introduced and studied in the context of code CFTs in \cite{Henriksson:2022dml}.

We note that the expression in \eqref{numberofbinarycodes} assumes the underlying codes (TBCs) exist, which in particular requires $n-\bar n=0\,\, {\rm mod}\,\, 8$. Applying this formula for $n,\bar n$ when this condition is not satisfied will lead to meaningless results.  The same applies to other expressions for $\NC$ found in this paper. 

When $p=2$, the Frobenius-Schur exponent is often twice larger than the exponent of $\mathscr{D}$, which in this case is $2$. 
To illustrate this  we consider $\Un(1)_2 \times \Un(1)_{-2}$, which admits a unique TBC -- the diagonal invariant with $(c_1,c_1)\in \C\subset \mathscr{D}=\Z_2^2$. 
One can define the  quotient $\mathscr{G}=\mathscr{D}/\C=\Z_2$ that consists of the elements of the form $(c_1,0)\in \mathscr{G}$ 
so that  $\mathscr{D}=\mathscr{G} \oplus \C$. The basis  $\kket{(a,b)}$ defined in \eqref{newbasis} with $a,b,\in\Z_2$ is well-defined and non-degenerate but has to be extended to $a,b\in \Z_k$ with $k=4$ because  $\kket{(a+2a_1,b+2b_1)}=e^{i\pi(a b_1+a_1 b)}\kket{(a,b)}$ is in general different from $\kket{(a,b)}$. 

\subsection{$k=4$ theories}
There are four basic theories with $k=4$: the level-$4$ toric code, $\Un(1)_2$ and $\Un(1)_{-2}$, and
an analog of \eqref{SO8} defined by quadratic form 
\bea
\label{qtheory}
q(c)=(\alpha^2+\alpha\beta+\beta^2)/4 ,\quad c=(\alpha,\beta)\in \Z_4^2.
\eea
This theory has vanishing chiral central charge and admits a unique topological boundary condition. To evaluate $\N$ for $n$ copies of this theory we need to express 
\bea
f(X)={1\over 2^{2h}}\sum_{\alpha,\beta\in \Z_2^g} e^{i\pi (\alpha^T X \beta+\alpha^T X \alpha+\beta^T X \beta)}
\eea
in terms of the invariants of $X$, which is a symmetric binary matrix. We can bring $X$ to canonical form \eqref{Xmat}, with $D_0$ being either identity matrix or $J$. In the latter case $f(X)=2^{-r}$, where $r$ is rank of $X$. In the former case $f(X)=(-2)^r$. Combining all together we find  
\bea
 \nonumber
 \NC&=&\sum_{h=0}^\infty \mu(h,p)\, 2^{2hn-h(h+1)/2} \sum_{r=0}^h  N(h,r,2)\, (-1)^{nr}2^{-nr},
\eea
which differs from the $k=4$ toric code result \eqref{toricp2} only by the factor $(-1)^{nr}$. Thus, $\NC$ would be the same as for $n$ copies of the toric code when $n$ is even, but would differ when $n$ is odd. For $n=1,2,3,4,5\dots$ we find $\NC=1,22,18,36766,2261326,\dots$.

\subsection{$k=2^m$ theories}
Theories with $k=8$, besides the level-8 toric code and $\Un(1)_{\pm 4}$, include theories with quadratic forms $q(c)=\pm 5 c^2/8$, $c\in \Z_{k/2}$, the analog of theories \eqref{SO8} and \eqref{qtheory}, and their combinations. This pattern continues for all $k=2^m$, $m\geq 3$; namely, there are six basic theories that can be combined \cite{Cano:2013ooa}. We do not attempt to consider all of these cases systematically, but instead sketch the necessary steps to do so.

The key task is to express $f(X)$ in \eqref{fXdef} in terms of the invariants of $X$. There are two ways to do this.
A more systematic approach is to bring the matrix $X$, with values in $\Z_{k/2}$, to a canonical form and then evaluate the sum over $c_i\in \mathscr{D}$. The canonical form of a symmetric matrix with values in $\Z_{2^{m-1}}$ was studied by Conway and Sloane \cite{conway2013sphere,allcock2015conway}. Their result is that $X$ can always be block-diagonalized into a direct sum of a diagonal matrix and a set of $2\times 2$ blocks of the form $2^e U$ or $2^e V$, where $e<m-1$ and
\bea
U = \begin{pmatrix}
0 & 1 \\
1 & 0
\end{pmatrix},
\qquad
V = \begin{pmatrix}
2 & 1 \\
1 & 2
\end{pmatrix}.
\eea
A full characterization of the invariants of $X$, which is necessary to derive the analogs of \eqref{fXgeneralT} and \eqref{GgeneralT} for $k=2^m$, is encoded in the Conway–Sloane normal form and is rather involved. We therefore leave the task of working out the explicit form of $f(X)$ and $\G_h$ for general $k=2^m$ theory to future work.

A more pragmatic approach is to represent $X = X_0 + 2 X_1 + \dots$ and bring the matrices $X_i$ into canonical form one by one, similarly to the approach we have taken in \eqref{Xmat}. We do not have a proof that this approach is sufficient for a general $k=2^m$ theory, but it works in practice, as illustrated by the examples of $\Un(1)_4$ and $\Un(1)_8$ discussed below.

\subsection{$\Un(1)_4$}
\label{U14}
We now proceed to consider  $\Un(1)^{n}_4 \times \Un(1)^{\bar n}_{-4}$ CS theory,  
which has  $k=8$. This example is similar to the case of $\Un(1)_2$ considered above. The Lagrangian subgroups $\C$ can be interpreted as type II codes over $\Z_4$ -- self-dual codes with the (modified if $\bar n\neq 0$) Hamming weight of all codewords divisible by $8$ \cite{CONWAY199330,568705}. To calculate their number  
we need to evaluate 
\bea
\label{fU4}
f(X)={1\over 4^h}\sum_{c\in \Z_{4}^h} e^{i\pi c^T X c/2}.
\eea
First, we represent $X=X_0+2X_1$ and bring $X_0$ to  block-diagonal form.  There are two cases to consider: (i) the non-degenerate block of $X_0$ can be brought to be the identity matrix of size $r$; (ii) the non-degenerate block of $X_0$ is the symplectic form $J$. 
In case (i), for the first $r$ variables $c_i$ we have the following sum,
\bea
{1\over 2}\sum_{c_i \in \Z_2} e^{i\pi c_i^2(1/2+(X_1)^{ii})}=2^{-1/2} e^{(2(X_1)^{ii}-1)i\pi/4}. 
\eea
The remaining $h-r$ variables $c_i$ give the same result as in \eqref{Gval2}.  For the theory $\Un(1)^{n}_4 \times \Un(1)^{\bar n}_{-4}$ to admit topological boundary conditions $n-\bar n$ should be divisible by $8$, hence in the resulting theory the phase factor $e^{(2(X_1)^{ii}-1)(n-\bar n)i\pi/4}$ will become trivial. 

For case (ii), the sum to calculate is 
\bea
{1\over 4}\sum_{a_1,a_2\in \Z_2} e^{i\pi a_1 a_2 +(X_1)^{11} a_1^2+(X_1)^{22} a_2^2}=2^{-1} e^{i\pi (X_1)^{11}(X_1)^{22}}.
\eea
And again when $n-\bar n=0\,\, {\rm mod}\,\, 8$ the phase factor is trivial. Combining the two cases we find 
\bea
\G_h= 4^{-h(h+1)/2}\sum_X f(X)=\sum_{r=0}^h N(h,r,2) 2^{-h(h+3)/2+r} 2^{-(n+\bar n)r/2},
\eea
and 
\bea 
\NC(\Un(1)^{n}_4 \times \Un(1)^{\bar n}_{-4})=\sum_{h=0}^\infty {2^{-h(h+2)+(n+\bar n)h} \sum_{r=0}^h N(h,r,2) 2^{r(1-(n+\bar n)/2)}\over \prod_{i=1}^\infty (1+2^{-i}) \prod_{i=1}^h (1-2^{-i})}.
\eea
This expression should admit a representation similar to \eqref{toricp2} \cite{gaborit1996mass,568705}.
For $n=\bar n=1,2,3,\dots$ we find $\NC=2,10,134,5662,\dots$

\subsection{$\Un(1)_8$}
Our final example is the case of $\Un(1)^{n}_8 \times \Un(1)^{\bar n}_{-8}$ CS theory, which has $k=16$.
We need to evaluate 
\bea
\label{fU8}
f(X)={1\over 8^h}\sum_{c\in \Z_8^h} e^{i \pi c^T (X_0+2X_1+4X_2) c/4}.
\eea
Upon diagonalizing $X_0$ and expanding $c=c_0+2c_1+4c_2$ with each $c_0,c_1,c_2\in \Z_2^h$ we find that the summand does not depend on $c_2$ and $c_1$ enters through the combination 
\bea
e^{i\pi c_1^T (X_0)(1+c_0)}
\eea
where we used that for a binary variable $c_1=c_1^2$. Now, we diagonalize $X_0$ as before. If the non-degenerate block of $X_0$ is identity, this forces first $r_0$ components of $c_0$ to be equal to one. If instead the non-degenerate block of $X_0$ is $J$, then first components of $c_0$ have to be zero. In any case the expressions involving first $r$ components of $a_0$ become one (where we used that $n-\bar n$ is divisible by 8). This gives the following factor $2^{-(n+\bar n) r}$. The remaning calculation is the same as for $\Un(1)_4$, leading to 
\bea 
&&\NC(\Un(1)^{n}_8 \times \Un(1)^{\bar n}_{-8})=
\sum_{h=0}^\infty {2^{-3h(h+1)/2+3/2(n+\bar n)h} \over \prod_{i=1}^\infty (1+2^{-i}) \prod_{i=1}^h (1-2^{-i})} \times \\ \nonumber
&& \sum\limits_{r_0=0}^h N(h,r_0,2) 2^{r_0(r_0+1)/2+r_0(h-r_0)-r_0 (n+\bar n)} \sum\limits_{r_1=0}^{h-r_0} N(h-r_0,r_1,2)2^{(r_0+r_1-h)-r_1 (n+\bar n)/2}.
\eea 
For $n={\bar n}=1,2,3,\dots$ we find $\NC=3,34,1926,\dots$

\section{Non-Abelian TQFTs}
\label{nonAbelian}
The derivation of the mass formula outlined in section \ref{derivation} and a more general construction of  \cite{Dymarsky:2024frx}  are valid in the non-Abelian case, modulo an assumption that the TBC states $|\C\rangle$ (that are in one to one correspondence with the Lagrangian algebras of $\T$) span the space of modular invariants in $\H^g$ when $g$ is sufficiently large. This assumption can be recast as an expectation that the modular bootstrap constraints in the limit of infinite genus are sufficiently powerful  to identify all rational CFTs with a given chiral algebra. Repeating the derivation of holographic duality between a non-Abelian TQFT gravity and the ensemble of boundary CFTs, one finds the ensemble weights to be inversely proportional to the norm of the TBC states evaluated at infinite genus \cite{Dymarsky:2024frx}. In remarkable progress, it was shown in  \cite{Barbar:2025vvf} that  in the limit of large $g$,   the states $|\C\rangle$ become mutually orthogonal and
$\langle \C|\C\rangle=\D^{g-1} |{\rm Aut}(\C)|$, where $|{\rm Aut}(\C)|$ 
understood in the SymTFT picture as the number of topological lines living on the boundary corresponding to  $\C$. 
With this simplification the mass formula takes the form 
\bea
\label{nonAmf}
\M(\T)=\sum_\C {\D\over |{\rm Aut}(\C)|}=\lim_{g\rightarrow \infty} {\D^{g}\over | {\rm MCG}(\Sigma_g)|} \sum_{\gamma\in {\rm MCG}(\Sigma_g)} {}^g\langle 0|U_\gamma|0\rangle^g.
\eea
The Abelian case is recovered by taking into account that $|{\rm Aut}(\C)|=\D$ for all $\C$.

To interpret the RHS as an average  over all connected 3d manifolds, a more natural normalization would be to redefine the mass $\M \rightarrow \M/\D$. We choose to  work with current normalization that gives the total number of TBCs in the case of an Abelian theory. 

In the general  non-Abelian case, the orbit of the mapping class group acting on $\H^g$ (for $g>1$) is infinite; hence the average in \eqref{nonAmf} is defined only formally. When the Hilbert space $\H^g$ is finite-dimensional, the  sum in \eqref{nonAmf} is nevertheless well-defined because it is a projection of $|0\rangle^{g}$ onto the modular-invariant subspace of $\H^g$.


There is also a similar expression for the number of TBCs,
\bea
\label{NCnA}
\NC(\T)=\lim_{g\rightarrow \infty} {1\over |{\rm MCG}(\Sigma_g)|} \sum_{\gamma\in {\rm MCG}(\Sigma_g)} {\rm Tr}_{\H^g}\, (U_\gamma).
\eea
For non-Abelian theories $\M$ and $\NC$ are different. In the context of TQFT gravity, the former is the normalization factor necessary for the weights of boundary ensemble to sum to one, while the latter is the dimension of the Hilbert space of the baby universe. 


The definition of the mass  as a weighted sum over 2d CFTs  sharing the same chiral algebra was earlier introduced in \cite{Hoehn} in the context of framed Vertex Operator Algebras (VOA). We review and reproduce their result from the sum over topologies below. 

\subsection{${\rm Vir}_{c=1/2}$}
Let us consider the 3d Ising  category -- the three-dimensional TQFT evaluating conformal blocks of $c=1/2$ Virasoro algebra ${\rm Vir}_{c=1/2}$. As in the Abelian case we will be considering  $n$ copies of the Ising TQFT and $\bar n$ copies of its conjugate and denote this theory by $\T_{n,\bar n}$.

In such a theory  a topological boundary condition would exist whenever $n-\bar n$ is divisible by $16$. A given TBC defines a 2d CFT with central charges $c_L=n/2,c_R=\bar n/2$ and chiral algebra ${\rm Vir}_{c=1/2}^{\otimes n} \times  \overline{{\rm Vir}}_{c=1/2}^{\otimes \bar n}$.  Focusing on the chiral case $\bar n=0$,  by analyzing corresponding VOAs  \cite{Hoehn} has found for the mass  (in our normalization), c.f.~\eqref{toricp2} and \eqref{symplp2},
\bea
\label{Isingmass}
\M(\T)=\sum_\C {\D\over |{\rm Aut}(\C)|}=\sum_{k\geq 1} 2^{k(k-1)/2+1} \sigma(k).
\eea
Here $\sigma(k)$ is the number of so-called D-codes -- linear, binary, triply-even codes $[n,k]$ of length $n$ and with $k$ generators, that include $c_{\vec{1}}=(\underbrace{1,\dots,1}_n)$ as one of the codewords.  The appearance of D-codes  is not entirely surprising given that any  framed VOA (a 2d CFT with $c_L=n/2,c_R=0$ and chiral algebra ${\rm Vir}_{c=1/2}^{\otimes n}$) defines such a code \cite{dong1998framed}. 

By considering non-zero $\bar n$ we slightly generalize \eqref{Isingmass} to define $\sigma(k)$ as the total number of  linear, binary codes of type $[n+\bar n,k]$, i.e.\ of length $n+\bar n$ with exactly $k$ generators, that include the codeword 
\bea
c_{\vec{1}}=(\underbrace{1,\dots,1}_n|\underbrace{1,\dots,1}_{\bar n}),
\eea  
and that are triply even with respect to the modified Hamming weight \eqref{modifiedweight},
\bea
\label{mhw}
w(c)=\sum_{1}^n c_i -\sum_{n+1}^{n+\bar n} c_i,\qquad c=(c_1,\dots,c_{n}|c_{n+1},\dots,c_{n+\bar n}).
\eea
A justification of \eqref{Isingmass} for the $n,\bar n\neq 0$ case on the field theory side will be given elsewhere. Below we  derive this result from \eqref{nonAmf} by summing over topologies.

The Hilbert space $\H^g$ of the Ising TQFT is spanned by conformal blocks that can be written explicitly in terms of theta functions \cite{Jian:2019ubz},
\bea
\label{basisI}
 \kket{(a,b)}={\vartheta^{1/2} {\genfrac{[}{]}{0pt}{}{b/2}{a/2}}(\Omega|0) / \Phi(\Omega)}/2^{g/2}.
\eea
Here  $a,b$ are binary strings of length $g$, $\Omega$ is the modular parameter of a Riemann surface $\Sigma_g$, and $\Phi(\Omega)$ is some $a,b$-independent factor.
These states are canonically normalized and the genus-$g$ vacuum state is given by, c.f.~\eqref{0},
\bea
\label{Ising}
|0\rangle^g={1\over \D^{g/2}}\sum_{a \in \Z_2^g} \kket{(a,0)},\qquad \D=2. 
\eea
As in the case of the toric code, there is also an analog of the basis  $|(\alpha,\beta)\rangle$ related to \eqref{basisI} by the Fourier transform of $a$. It will be discussed elsewhere. 

The action $U$ of the mapping class group  on  $\H^g$ is discussed in \cite{Gilmer,  Jian:2019ubz}. There is an exact sequence 
\bea
\label{sequence}
1\rightarrow U(H_g)  \rightarrow  U({\rm MCG}(\Sigma_g)) \xrightarrow{\rho} \Sp(2g,\Z_2)\rightarrow 1, 
\eea
where $H_g$ is the subgroup of ${\rm MCG}(\Sigma_g)$ that acts trivially on $\Z_2$-valued cycles in $H^1(\Sigma_g,\Z_2)$.
 Modulo phase factors, the group $\Sp(2g,\Z_2)$
acts on  the basis \eqref{basisI}  by an affine action on the labels, 
\bea
(a,b)\rightarrow \gamma\, (a,b)+v(\gamma),\qquad \gamma\in \Sp(2g,\Z_2),
\eea
where $v(\gamma)$ is a binary vector of length $2g$ that only depends on $\gamma$, and all algebra is understood mod $2$. 
The kernel of $\rho$, which is a normal subgroup of $U({\rm MCG}(\Sigma_g))$, acts on the basis elements \eqref{basisI} by a pure phase. This action is crucial for what follows.  

Using the same logic as in Appendix \ref{measureApp},  the exact sequence \eqref{sequence}  can be used to  rewrite the average  over $U({\rm MCG}(\Sigma_g))$ in \eqref{nonAmf} as two averages, over ${\rm Ker}(\rho)$ and $\Sp(2g,\Z_2)$.
As a result   the mass can be written as follows
\bea
\M=\sum_{a=0}^\infty \mu(a,2)\, \D^a \G_a,\qquad \D=2^{n+\bar n},
\eea
where 
\bea
\label{GaI}
\D^h \G_h ={\D^h \over |{\rm Ker}(\rho)|} \sum_{U\in {\rm Ker}(\rho)} {}^h \langle 0|U|0\rangle^h.
\eea
To derive this we used that $U_{\gamma_h}$ for  $\gamma_h= {\bf e}^h \oplus {\bf s}^{g-h}$ acts on $ \kket{(a,b)}$ by simply exchanging $a_i\leftrightarrow b_i$ for $h<i\leq g$, while elements of $ {\rm Ker}(\rho)$ act by a pure phase. 

In the expression above $|0\rangle^h$  stands for the vacuum state of genus $h$ of the full theory 
$\T_{n,\bar n}$. Given that elements of ${\rm Ker}(\rho)$ act on $ \kket{(a,0)}$ by phases, $\D^h \G_h$ is then given by the total number of states of the form 
\bea
\label{string}
 \kket{(a_1,0)} \otimes \dots \otimes  \kket{(a_n,0)}\otimes \overline{ \kket{(a_{n+1}, 0)}}\otimes \dots \otimes \overline{ \kket{(a_{n+\bar n}, 0)}}, \quad a_i \in \Z_2^g,
\eea
invariant under all phase factors.  The way ${\rm Ker}(\rho)$ acts on the state \eqref{string}, and the orbit of this state under the action of $\Sp(2g,\Z_2)$, both depend on the number of linearly-independent differences $a_i-a_j$ mod $2$. Let us assume there are $k$ linearly independent differences; we then can introduce a basis in the space of differences,  $e_j \in \Z_2^g$, with $j=1,\dots, k$, such that 
\bea
\label{ajrep}
a_i=a_1+\sum_{j=1}^k e_j\, G_i^j, 
\eea
where $G_i \in \Z_2^k$ and all algebra is mod $2$. Choosing a basis allows us to rewrite $a_i$, which are binary strings of length $g$,  as binary strings of length $k$ that we denote $G_i$.  The state \eqref{string} is therefore parametrized by a $k \times (n+\bar n)$ binary matrix $G$.  There are $2^{k+1}$ generators of  ${\rm Ker}(\rho)$, each acting  \eqref{string}  by a phase 
\bea
\label{phases}
e^{i \pi w(c)\over 8},
\eea  
where $c\in \Z_2^{n+\bar n}$ is an arbitrary linear combination (mod $2$) of the rows of matrix $G$, extended by an extra row of identities $c_{\vec{1}}$, and $w(c)$ is the modified Hamming weight defined in \eqref{mhw}.  The appearance of the extra row consisting of $n+\bar n$ identities is easy to understand: multiplying all states $ \kket{(a_i,0)}$  in \eqref{string} by the same phase $e^{i\pi/4}$ is a symmetry because $n+\bar n$ is divisible by $16$ (and hence also by 8). 
Thinking of the extended $(k+1)\times (n+\bar n)$ matrix as a generating matrix of a binary $[n+\bar n,k+1]$ code, the triviality  of all of the phases \eqref{phases} implies that the code is triply-even, i.e.~the modified Hamming weight of all codewords is divisible by $8$. 

It is straightforward to see that choosing a different basis of differences $e_j$ or choosing a different ``reference element'' in \eqref{ajrep}, e.g.~$a_j=a_2+\sum_{j=1}^k e_j G_i^j$, would lead to an equivalent matrix $G$ generating the same code.  Considering an equivalent generating matrix $G$ with the same reference element and differences $e_j$ would lead to a different state \eqref{string}, also invariant under ${\rm Ker}(\rho)$. To evaluate the total number of states that correspond to the same code we note that $a_1$ can be choosen as any of $2^g$ states, $a_1+e_1$ is any of $2^g-1$ states, $a_1+e_2$ is any of $2^g-2$ states etc. We thereofre find for \eqref{GaI}
\bea
\label{GaIsing}
\D^a \G_a = \sum_{k\geq 0} 2^g \prod_{i=0}^{k-1} (2^g-2^i) \sigma(k+1).
\eea
There is no known closed-form expression for the number of triply-even D-codes with $k$ generators $\sigma(k)$, but for small $k$ and $n+\bar n$ it can be calculated using computer algebra and the following representation 
\bea
\nonumber
\sigma(k+1)&=&{1\over K\prod_{i=0}^{k-1}(K-2^i)}\sum_{\substack{n_0+\cdots+n_{K-1} = n,\\
n_0,\dots,n_{K-1} \ge 0 }} \sum_{\substack{{\bar n}_1+\cdots+{\bar n}_{K-1} = {\bar n},\\
{\bar n}_0,\dots,{\bar n}_{K-1} \ge 0 }} {n!\over n_0 !\dots n_{K-1} !} {{\bar n}!\over {\bar n}_0 !\dots {\bar n}_{K-1} !} f(n_\alpha,\bar n_\alpha),\\
f(n_\alpha,\bar n_\alpha)&=&\begin{cases}
1, &
\begin{aligned}
&\text{rank}({\sf G})=k+1, \\ \nonumber
&(n_0-\bar n_0,\dots,n_{K-1}-\bar n_{K-1})\cdot  c = 0\,\,{\rm mod}\,\, 8,\,\, \text{for any}\,\, c=v^T G \,\,{\rm mod}\,\, 2,\, v\in \Z_2^{k+1},
\end{aligned}
\\[6pt]
0, & \text{otherwise.}
\end{cases}
\eea
This formula requires a few clarifications. Here  $K=2^k$, while matrix $G$ is  $(k+1) \times K$ consisting of the binary representations of $\alpha=0,\dots, K-1$, extened by the codeword of  identities of length $K$.  Finally, matrix  ${\sf G}$ includes only the columns of $G$ for which $n_\alpha+\bar n_\alpha>0$. Its $\text{rank}({\sf G})$ should be evaluated in algebra over $\Z_2$ (over the Galois field $F_2$ to be precise). 
The derivation, and interpretation, of this formula is straightforward:  $n_\alpha,{\bar n}_\alpha$ are the number of times $a_1+e_j \alpha^j$, where $\alpha^j$ is the binary representation of $\alpha$, appears in the sets $\{a_1,\dots, a_{n}\}$ and $\{a_{n+1},\dots, a_{n+\bar n}\}$ correspondingly. 

It is interesting to note that so far $g(g+1)/2\geq 2^g-1$, i.e.~$g=1,2$,  and for arbitrary $n,\bar n$ \eqref{GaIsing} coincides with the naive expression, c.f.~\eqref{fU8},
\bea
\label{naive}
\G_a={1\over 8^{g(g+1)/2}} \sum_X f(X),\quad  f(X)={1\over 2^g}\sum_{c\in \Z_2^g} e^{i \pi c^T X c/4},
\eea
where the sum is over symmetric $g\times g$ matrices $X$ with values in $\Z_8$. This expression follows from the action of the mapping class group on $U(1)_1$ characters $\vartheta {\genfrac{[}{]}{0pt}{}{b/2}{a/2}}(\Omega|0)$, which naively suggests that $U({\rm MCG}(\Sigma_g))$ acts on the vacuum state as $\Sp(2g,\Z_{16})$.  
For $g\geq 3$ the number of phase generators \eqref{phases} exceeds that one in $\Sp(2g,\Z_{16})$, indicating this is incorrect.  

Going back to mass formula, using \eqref{toriccodeN} we readily find
\bea
\sum_{a=0}^\infty \mu(a,2)2^a \prod_{i=0}^{k-1} (2^a-2^i)=2^{k(k+1)/2+1},
\eea
thus reproducing \eqref{Isingmass}. For small $n=\bar n=1,2,3,4,5,6,\dots$ we find 
\bea
\label{IsingTBC}
\NC(\T_{n=\bar n})=2,10,134,4746,419862, 88276726,\dots
\eea
which is in agreement with the brute force calculation by summing over all topological boundary conditions constructed via condensation into $n$ copies of the toric code, as discussed in the End Matter of \cite{Barbar:2023ncl}. For $n=\bar n=1,2,3$ \eqref{Isingmass} is the same as for $U(1)^n_4 \times U(1)^{\bar n}_{-4}$ theory  because for small $n=\bar n\leq 3$ \eqref{GaIsing} and the average over \eqref{fU4} are the same. 
For chiral theories with $n=16,\bar n=0$ we find $\NC(\T_{16})=140668954142$ in agreement with the results of \cite{griess2001virasoro}, which calculated $\sigma(k)$ for that case.

\section{Five-dimensional Abelian $2$-form theory } 
\label{sec:5d}
The notion of mass and the approach to calculating it by summing over all connected topologies can be extended to other dimensions. 
Here we discuss the case of 5d Abelian BF-type theory of $2$-forms, specified by the action 
\bea
\label{5dtheory}
{K_{ij}\over 4\pi}\int B^i \wedge dB^j,
\eea
where $K_{ij}$ is a $g\times g$ antisymmetric integer-valued matrix and $B^i$ are two-forms. This theory is closely related to 3d bosonic Chern-Simons. Similarly to the 3d case where the $K$-matrix is the Gram matrix of a lattice $\Lambda$, in 5d $K_{ij}$ is the Gram matrix of a symplectic lattice $\Lambda\subset \R^{g}$, i.e.~an integer-valued skew-symmetric form. Similarly to 3d, the surface operators of this theory are labeled by elements of the discriminant group $c\in \mathscr{D}=\Lambda^*/\Lambda$, and (up to an identification discussed below) topological boundary conditions are specified by Lagrangian subgroups $\S\subset \mathscr{D}$ -- maximal non-anomalous subgroups of elements $c\in \S$ with trivial mutual braiding defined in terms of $K$.  These subgroups have size $\D=| \mathscr{D}|^{1/2}$, which must be an integer in order for TBCs to exist. 

Similarly to 3d where Lagrangian subgroups $\C$ can be interpreted as classical even self-dual codes, in 5d Lagrangian subgroups $\S$ can be interpreted as  classical symplectic self-dual codes (maximal isotropic subspaces) \cite{Barbar:2025krh}.  The latter define quantum stabilizer codes on $g$ qudits, leading to another interpretation of $\S$ \cite{Barbar:2025krh,Radhakrishnan:2025gzl}.

Any integer-valued antisymmetric $K_{ij}$ can be brought to canonical form, when the theory \eqref{5dtheory} decomposes into a product of theories  $\T_k$ with action 
\bea
\label{5d}
{k\over 4\pi}\int B \wedge dC,
\eea
familiar from the low-energy limit of IIB String Theory on AdS${}_5\times \mathbb{S}^5$ \cite{Witten:1998wy}.  These theories, and their combinations, are known to describe the global symmetries of 4d gauge theories. In particular, the Lagrangian subgroups of the theory consisting of  $g$ copies of \eqref{5d} would correspond to different 4d YM theories with gauge algebra $su(k)^g$ \cite{aharony2013reading,gaiotto2015generalized}.

The theory consisting of $g$ copies of \eqref{5d} on a signature-zero four-manifold whose  second cohomology group has size $|H_2(M_4,\Z)|=2n$ is closely related to the 3d theory consisting of  $n$ copies of the level-$k$ toric code on a Riemann surface of genus $g$. This observation can be used to quantize the 5d theory and define the analog of the basis \eqref{newbasis} that makes manifest the action of the mapping class group of the four-manifold, which is a subgroup of $\Or(n,n,\Z)$. Similarly to \eqref{modular}, $\gamma \in \Or(n,n,\Z)$ acts by the fundamental action of $\gamma\,\,{\rm mod}\,\, k$ on a vector in $\Z_k^{2g}$. Furthermore, the Lagrangian subgroups $\S$ of this 5d theory correspond to symplectic codes $\S$ of the 3d toric code, justifying the use of the same notation for both \cite{Barbar:2025krh}.

The holographic duality between $g$ copies of the 5d theory \eqref{5d} summed over topologies and the boundary ensemble specified by the Lagrangian subgroups $\S$ readily leads to a mass formula generalizing \eqref{mass},
\bea
\label{5dmass}
{\mathcal N}(\T_k^{\otimes g})=\sum_\S 1= \lim_{n\rightarrow \infty} {\D^n\over |\Or(n,n,\Z_k)|} \sum_{\gamma \in \Or(n,n,\Z_k)} {}^n\langle 0|U_\gamma|0\rangle^n.
\eea
Here $|0\rangle^n$ is the vacuum state, the TQFT path integral on a 5d  ``handlebody,'' a manifold bordant to $\#^n(\mathbb{S}^2\times \B^3)$. Thus, modulo subtleties discussed in \cite{Barbar:2025krh}, the RHS can be interpretation as a sum over closed connected 5d manifolds, giving rise to  renormalized partition function of TQFT gravity. These subtleties are unimportant to justify \eqref{5dmass}, which only relies on the fact that the states $|\S\rangle$ defined by Lagrangian subgroups (self-dual symplectic codes) $\S$ span the space of all invariants under the action of $\Or(n,n,\Z_k)$. 

This logic suggests that $\NC$ counts the number of TBCs of the Abelian 5d theory \eqref{5dtheory}, identified by the state these boundary conditions define in terms of the sandwich construction. Namely, two TBCs  are considered equivalent if corresponding states are the same up to normalization.

To evaluate $\NC$ we first note that, 
similarly  to \eqref{product} factorization of $\Or(n,n,\Z_k)$ for composite $k$ readily yields 
\bea
\label{product5d}
{\mathcal N}(\T^{\otimes g}_k)=\prod_i \NC(\T^{\otimes g}_{k_i}),\quad k=\prod_i k_i,
\eea
where $k_i=p_i^{m_i}$ are powers of  prime $p_i$.  In what follows we therefore can assume that $k=p^m$ for a prime $p$. 

The vacuum state $|0\rangle^n$ is invariant under the subgroup $\Gamma_0(n,n,\Z_k)\subset \Or(n,n,\Z_k)$. As in the case of \eqref{1X}, using similarity transformations from $\Gamma_0$ any $\gamma\in \Or(n,n,\Z_k)$ can be brought to the form 
\bea
\gamma_h(X)= {\bf e}(X) \oplus {\bf s}^{n-h},\quad 
{\bf e}(X)=\left(
\begin{array}{cc}
 1_h & 0 \\
 p\, X & 1_h \\
\end{array}
\right), \quad {\bf s}^{n-h}=\left(
\begin{array}{cc}
0 & 1_{n-h} \\
 1_{n-h} & 0 \\
\end{array}
\right),
\eea
where $X$ is antisymmetric. This means $X_{ii}=0$, including the case of $p=2$, see Appendix \eqref{orthogonal}. 
We now introduce the measure, c.f.~\eqref{measure},
\bea
\label{nu}
\nu(h,n,p)={|\Gamma_0(n,n,\Z_k)|^2 p^{(m-1)h(h-1)/2} \over |\Or(n,n,\Z_k)| |S_h|}=\qbinom{g}{h} {p^{(n-h)(n-h-1)/2} \over \prod_{i=0}^{n-1} (p^i+1)}, 
\eea
where $S_h$ is the subgroup of $\Gamma_0$ that preserves the form of  $\gamma_h(X)$, 
\bea
|S_h|=|\GL(h,\Z_k)|  |\GL(n-h,\Z_k)| k^{h(n-h)} \times k^{h(h-1)/2} \times k^{h(n-h)}.
\eea
Similarly to the 3d case, the measure \eqref{nu} is  $m$-independent.  It further simplifies in the  $n\rightarrow \infty$ limit, c.f.~\eqref{muinf},
\bea
\nu(h,p)=\lim_{n\rightarrow \infty}\nu(h,n,p) ={p^{-h(h-1)/2}\over 
 \prod_{i=0}^\infty (1+p^{-i}) \prod_{i=1}^h (1-p^{-i})}.
\eea
Now the mass formula  reads 
\bea
{\mathcal N}(\T^{\otimes g}_k) =\sum_{h=0}^\infty \nu(h,p) \D^{h} \FF_{h},\qquad \D=k^g,
\eea 
where we introduced 
\bea
\FF_{h}=p^{(1-m){h}(h-1)/2}\sum_X \left(f(X)\right)^g.
\eea
Here 
\bea
\label{5df}
f(X)={}^{h}\langle0|U_\gamma|0\rangle^{h}={1\over k^{2{h}}} \sum_{a,b \in \Z_k^{\tilde a}} e^{{2  \pi i p \over k} a^T X b},\quad k=p^m.
\eea
This expression is easy to calculate by representing $X=X_0+pX_1+\dots$ and then iteratively bringing $X_i$ to the form \eqref{Xmat}, where 
each $D_{r_i}$ is  the canonical symplectic form of size $r_i$
\bea
D_{r_i}=\left(\begin{array}{cc}
0 & -1_{r_i/2}\\
1_{r_i/2} & 0\end{array}\right). 
\eea
The number of such rank $r$ (which must be even) matrices of size $h$ is given by  
\bea
 \label{NarlO}
N(h,r,p)&=& {|\GL(h,\Z_p)|\over |\Sp(r,\Z_p)| |\GL(h-r,\Z_p)| p^{r(h-r)}}=\\
&&
{\prod_{i=1}^h (1-p^{-i})\over  \prod_{i=1}^{h-r} (1-p^{-i})} p^{h r-r(r+1)/2}\prod_{i=0}^{r/2} \left(1-p^{-2i}\right).
\eea
We can extend this formula to assume that $N(h,r,p)=0$ when $r$ is odd. 

After bringing $X$ to the form of \eqref{Xmat}, the sum in \eqref{5df} over $a,b$ can be easily evaluated to be 
\bea
f(h,r_0,\dots, r_{m-2})=p^{- \sum_{i=0}^{m-2}(m-1-i)r_i},
\eea
leading to, c.f.~\eqref{tcN},
\bea
&& {\mathcal N}(\T_{k}^{\otimes g})=\sum_{h=0}^\infty \nu(h,p)\, p^{h m g}\, \FF_h,\\ \nonumber 
&& \FF_h= \sum_{r_0=0}^h  p^{-h(h-1)/2}N(h,r_0,p) \sum_{r_1=0}^{h-r_0} p^{-(h-r_0)(h-r_0-1)/2}N(h-r_0,r_1,p)\times \\  \nonumber 
&& \dots \times    \sum_{r_{m-2}=0}^{h-\tilde{r}}  p^{-(h-\tilde{r})(a-\tilde{r}-1)/2} N(h-\tilde{r},r_{m-2},p) \, p^{-n \sum_{i=0}^{m-2}(m-1-i)r_i}, \quad \tilde{r}=r_0+\dots +r_{m-3}.
\eea

When $k=p$, we readily find 
\bea
{\mathcal N}(\T_{p}^{\otimes g})&=&\sum_{a=0}^\infty {p^{h g-h(h-1)/2 }\over 
 \prod_{i=0}^\infty (1+p^{-i}) \prod_{i=1}^h (1-p^{-i})}\\
&=&\prod_{i=1}^g (p^i+1)= {|\Sp(2g,\Z_p)|\over |\Gamma_0(2g,\Z_p)|},
\eea
matching the size of the coset,  in agreement with \cite{Aharony:2023zit}.  

When $k=p^2$ we find, c.f.~\eqref{toricp2}, 
\bea
 \nonumber
{\mathcal N}(\T_{p^2}^{\otimes g})&=&\sum_{h=0}^\infty \nu(h,p)\, p^{2hg-h(h-1)} \sum_{r=0}^h  N(h,r,p)\, p^{-g r}=\\
 &&
\sum_{\tilde{g}=0}^g p^{\tilde{g}(\tilde{g}+1)/2} \sigma(\tilde{g}),\quad \sigma(\tilde{g})= 
 \prod_{i=1}^{{\tilde g}} {p^{2(\tilde g+1-i)}-1\over p^i-1}, 
\label{symplp2}
\eea
where $\sigma(\tilde{g})$ is the number of symplectic $[g,\tilde g]$ codes, of length $g$ and with exactly $\tilde g$ generators. This expression  matches the sum over the $g+1$ orbits of $\Sp(2g,\Z_{p^2})$ derived in \cite{Barbar:2025krh}.  

\section{Summary and Outlook}
\label{summary}
In this paper, we have introduced the notion of  mass  for  a three-dimensional TQFT as a count of topological boundary conditions, weighted by a symmetry factor  determined by the size of the associated symmetry group (see also \cite{Hoehn, Barbar:2025vvf}). In the holographic duality between TQFT gravity -- defined as the sum of the TQFT over all three-dimensional topologies -- and an ensemble of boundary CFTs, the mass is given by the TQFT partition function averaged over all connected closed three-dimensional manifolds obtained via Heegaard splittings. We use this prescription to evaluate the mass for several classes of TQFTs.

Focusing first on the Abelian case, where the mass reduces to a simple count of topological boundary conditions (TBCs), we show how to compute the mass for an arbitrary bosonic theory. We then define the mass in the non-Abelian case and evaluate it for $n+\bar n$ copies of the Ising modular tensor category. Finally, we extend the construction to five dimensions and  compute the number of the Lagrangian subgroups for any five-dimensional Abelian two-form theory.

On a conceptual level, the mass formula introduced here generalizes the classical mass formulas for codes, lattices, and theta functions \cite{conway2013sphere}. In particular, the case of certain self-dual codes arises naturally from the Abelian Chern–Simons theories studied in this work. The analogous result for even self-dual lattices -- the Smith–Minkowski–Siegel mass formula -- should conjecturally  emerge in the limit of infinite Chern–Simons level, a direction we leave for future investigation.

Our results further confirm that the gravitational bulk sum over all Heegaard splittings obtained via genus reduction  in \cite{Dymarsky:2024frx} is well defined and yields the expected results. On a practical level, the paper provides explicit and novel representations for the number of topological boundary conditions in bosonic Abelian theories and in multiple copies of the Ising TQFT. Analogous to how mass formulas in mathematics have been used to classify all codes or lattices of a given type \cite{PLESS1975313,CONWAY198026,conway2013sphere}, the known mass can be used to verify that the list of topological boundary conditions  for a given theory is complete.
We note that the explicit expressions derived in this paper for the number of TBCs implicitly assume the latter to exist. Using these expressions when there are no TBCs, e.g.~\eqref{masstypeII} with $n$ not divisible by $8$, would lead to meaningless results. 

In a parallel technical development, we extend the congruence property for bosonic Abelian theories by showing that the action of the mapping class group on the TQFT Hilbert space reduces to $\Sp(2g,\Z_k)$ for some $k$, with $\Gamma_0(2g,\Z_k)$ stabilizing the vacuum,  if and only if the theory admits topological boundary conditions. As a byproduct, we also provide a new derivation of the result of \cite{kaidi2022higher}, demonstrating that topological boundary conditions exist if and only if all higher central charges vanish (see Appendix~\ref{proof}). We also show that an Abelian bosonic theory $\T_p$, with  $p$ an odd prime, has a partition function on any connected closed 
3-manifold equal to that on a certain connected sum of lens spaces.


Our work raises a number of open questions. An immediate and largely technical problem is to complete the analysis for theories with the Frobenius-Schur exponent $k=2^m$ by finding closed-form expressions for $f(X)$ and $\G_h$. Another interesting question is to evaluate the number of TBCs starting from the expression \eqref{trH} which counts the dimension of the baby universe Hilbert space.
A more conceptual and challenging direction is to extend our study to  other non-Abelian theories, e.g.~$SU(2)_k$ Chern–Simons theory, for which the image of the mapping class group could be infinite. 
Ultimately, the ideas introduced in this paper may be applied to Virasoro TQFTs at large central charge, shedding new light on three-dimensional quantum gravity.

\acknowledgments
We thank Ahmed Barbar, Meng Cheng,  and Sahand Seifnashri for discussions. A.D.~acknowledges support from the IBM Einstein Fellow Fund and NSF under grant 2310426.

\appendix
\section{Size of $\Sa$}
\label{groups}
Throughout this section we assume $k=p^m$, where $p$ is prime. 
We will need the order of the group of invertible $g\times g$ matrices over $\Z_k$, 
\bea
|\GL(g,\Z_k)|&=&p^{(m-1)g^2}\prod_{i=0}^{g-1}(p^g-p^i)=p^{(m-1/2)g^2-g/2}\prod_{i=1}^g(p^i-1).
\eea
The group $\Sp(2g,\Z_k)$ is defined to be the group of matrices preserving skew-symmetric form 
\bea
J=\left(\begin{array}{cc}
0 & -1_g\\
1_g & 0\end{array}\right). 
\eea
Its order is 
\bea
|\Sp(2g,\Z_k)|&=&p^{(m-1)g(2g+1)+g^2}\prod_{i=1}^g (p^{2i}-1).
\eea
We then define the subgroup $\Gamma_0(2g,\Z_k)$ of all matrices of the form
\bea
\left(\begin{array}{cc} A & B \\
0 &D \end{array}\right)\in \Gamma_0(2g,\Z_k)\subset \Sp(2g,\Z_k).
\eea
This subgroup is the stabilizer of the vacuum state \eqref{0} that can be written as a code state $|\S_0\rangle$ defined by the symplectic code $(a_1,\dots ,a_g,0,\dots,0)\in \S_0$.
The size of this group can be easily evaluated,
\bea
|\Gamma_0(2g,\Z_k)|&=&|\GL(g,\Z_k)| p^{mg(g+1)/2}=p^{(3m-1)/2g^2+(m-1)g/2}\prod_{i=1}^g(p^i-1).
\eea

For  $\gamma_h(X)$ given by \eqref{1X}, which we now write explicitly as 
\bea
\gamma_h(X)=\left(\begin{array}{cc|cc}
1_h & 0 & 0 & 0\\
0 & 0_{g-h} & 0 & -1\\
\hline
pX & 0 & 1_h & 0\\
0 & 1 & 0 & 0_{g-h}
\end{array}\right),
\eea  
we define the subgroup  of matrices $\gamma_0\in S_a\subset \Gamma_0(2g,\Z_k)$ satisfying 
\bea
\gamma_h(X)\gamma_0=\tilde{\gamma}_0\gamma_h(\tilde{X})
\eea
for some $\tilde{\gamma}_0 \in  \Gamma_0(2g,\Z_k)$ and $\tilde{X}$. As a result we find that elements $S_a$ are arbitrary matrices of the form 
\bea
\label{Saelements}
\left(
\begin{array}{cccc}
a_{11} & a_{12} & b_{11} & b_{12} \\
 0 & a_{22} & b_{21} & 0 \\
 0 & 0 & d_{11} & 0 \\
 0 & 0 & d_{21} & d_{22} \\
\end{array}
\right)\in  \Gamma_0(2g,\Z_k),
\eea
while $\tilde{X}=a_{11}^T X a_{11}$.
To calculate the number of such matrices we need to take into account  that the block $A$ is any invertible $g\times g$ matrix and the block $A^{-1}B$ is symmetric and otherwise arbitrary. 
We therefore find 
\bea
|\Sa|=|\GL(h,\Z_k)|  |\GL(g-h,\Z_k)| k^{h(g-h)} \times k^{h(h+1)/2} \times k^{h(g-h)}.\\
\eea
This formula should be read as follows: number of all invertible $a_{11}$, number of all invertible $a_{22}$, number of all $a_{12}$, number of all symmetric $b_{11}$, number of all possible $b_{12}$ (we count matrices $B$ instead of $A^{-1}B$ since the number of possible matrices is the same). 

The final expression for $\mu(h,g,k)$ involves the Gaussian binomial coefficient, defined as follows
\bea
\qbinom{n}{k}
=
\prod_{i=0}^{k-1}\frac{p^{n-i}-1}{p^{k-i}-1}
=
\frac{(1-p^n)(1-p^{n-1})\cdots(1-p^{n-k+1})}{(1-p^k)(1-p^{k-1})\cdots(1-p)}. \label{binomial}
\eea

\section{Average over $\Sp(2g,\Z_{p^m})$ vs $\Sp(2g,\Z_{p})$}
\label{measureApp}
The average over the double-coset  $\coset_g=\Gamma_0(2g,Z_k) \backslash \Sp(2g,\Z_k) / \Gamma_0(2g,Z_k)$ can proceed in two ways. The first, discussed in the main text, is to choose representatives of the form $\gamma_a(X)= {\bf e}(X) \oplus {\bf s}^{g-a}$
 given in \eqref{1X}, and then sum over all representatives. An alternative route is to use the exact sequence 
 \bea
 1\rightarrow \Gamma_p^{(m)} \rightarrow \Sp(2g,\Z_{p^m}) \xrightarrow{\rho} \Sp(2g,\Z_{p}) \rightarrow 1, 
 \eea
 where $\Gamma_p^m$ is the principal congruence subgroup of level $p$ mod $p^m$, i.e.~the group of matrices of the form $\1+p A$ for some $A$ with elements in $\Z_{p^m}$.  The image of $\Gamma_p^{(m)}$ inside $\Sp(2g,\Z_{p^m})$ -- the kernel of map $\rho$ from $\Sp(2g,\Z_{p^m})$ to $\Sp(2g,\Z_p)$ --  is a normal subgroup of $\Sp(2g,\Z_{p^m})$. Therefore the average over $\Sp(2g,\Z_{p^m})$ can proceed into two steps (in arbitrary order): averaging over $\Sp(2g,\Z_{p})$, and averaging over the image of $\Gamma_p^{(m)}$ inside $\Sp(2g,\Z_{p^m})$,
 \bea
 {1\over |\Sp(2g,\Z_{p^m})|}\sum_{\gamma \in Sp(2g,\Z_{p^m})} {}^g\langle 0|U_\gamma|0\rangle^g= {1\over |\Sp(2g,\Z_{p})|}\sum_{\gamma \in Sp(2g,\Z_{p})}   = {}^g\langle 0|U_\gamma |\emptyset\rangle^g,\\
 |\emptyset\rangle^g={1\over |{\rm Ker}(\rho)|} \sum_{\gamma \in {\rm Ker}(\rho)} U_\gamma |0\rangle^g.
 \eea
The average over $\Sp(2g,\Z_{p})$ is in fact over the coset  $\Gamma_0(2g,\Z_p) \backslash \Sp(2g,\Z_p) / \Gamma_0(2g,\Z_p)$ with the representatives given by $\gamma_a ={\bf 1}^{a} \oplus {\bf s}^{g-a}$.   Using the representation \eqref{0} for the vacuum state and the explicit form of the modular action \eqref{modular}  we readily find 
\bea
{}^g\langle 0|U_{{\bf 1}^{h} \oplus {\bf s}^{g-h}} |\emptyset \rangle^g= {}^h\langle 0|\emptyset\rangle^h\, \D^{g-h}.
\eea
To evaluate the average over ${\rm Ker}(\rho)$  we note that the only elements that contribute to the average are of the form,  c.f.~\eqref{1X}, 
\bea
\left(
\begin{array}{cc}
 1_h & 0 \\
 p\, X & 1_h \\
\end{array}
\right) \in \Gamma_p^{(m)},
\eea
so that
\bea
{}^h\langle 0|\emptyset\rangle^h=p^{(1-m)h(h+1)/2} \sum_{X} {}^h\langle 0|U_{\gamma(X)}|0\rangle^h  =\G_h.
\eea

\section{Congruence property and higher central charges}
\label{proof}
We start with an Abelian theory $\T_p$ for a prime $p$ and the Frobenius-Schur exponent $k=p^m$ and {\it assume} that the vacuum state  $|0\rangle\in \H^1$ (for the theory on a torus) is stabilized by $\Gamma_0(k)\subset \SL(2,\Z)$. 
Then the average over $\SL(2,\Z)$ in \eqref{mainformula} can be written as follows
\bea
\D^{-1} \M_1=\D^{-1} +\sum_{n=0}^{k/p-1} \sum_{c\in \mathscr{D}} \theta_c^{p n}.
\eea
For each $c$ the sum over $n$ is either real positive or vanishes (this is the same statement as $\G_1\geq 0$, which we already made in the main text). Therefore the modular-invariant subspace of $\H^1$ is non-trivial, implying the theory admits TBCs.  A similar argument can be made for any $g$. We therefore conclude that in an Abelian theory $\Gamma_0(2g,\Z_k)\subset  \Sp(2g,\Z_k)$ stabilizes the vacuum state if and only if the theory admits  TBCs.

Now we relax the assumption that the vacuum state  $|0\rangle\in \H^1$ of $\T_p$ is invariant under $\Gamma_0(k)\in \SL(2,\Z)$, which necessarily means the theory admits no TBCs.
The congruence property \cite{Ng:2012ty} nevertheless guarantees that the vacuum  is invariant under $\Gamma_1(k)\in \SL(2,\Z)$.
We next consider the theory $\T_p \times \bar \T_p$. This theory always admits the diagonal invariant, hence its vacuum in $\H^1$ is stabilized by $\Gamma_0(k)\in \SL(2,\Z)$.  Any element from $\Gamma_0(2,Z_k)$ can be represented as an element from $\Gamma_1(2,\Z_k)$ multiplied (from the left or from the right) by a diagonal matrix
\bea
\gamma(x)=\left(\begin{array}{cc}
x & 0\\
0 & x^{-1}\end{array}\right)\in \SL(2,\Z_k),
\eea
with an  invertible $x\in \Z_k$. These elements act on the vacuum of $\T_p$ by a phase, which by assumption can not be trivial for all invertible $x$. Accordingly the average over $\SL(2,\Z)$ in \eqref{mainformula} vanishes, consistent with the expectation there are no modular invariant states in $\H^1$. 
Now we consider the partition function of $\T_p$ on the lens space $\bar{L}(n,1)$ for $n$  co-prime with $p$.  The corresponding element of $\SL(2,\Z_k)$ can be written as follows 
\bea
\gamma_n=\left(\begin{array}{cc}
1 & 0\\
n & 1\end{array}\right)=\gamma_1 \gamma(n) \left(\begin{array}{cc}
0 & -1\\
1 & 0\end{array}\right)  \tilde{\gamma}_1,
\eea
where $\gamma_1,\tilde{\gamma}_1\in \Gamma_1(2,\Z_k)$. From here it immediately follows that the phase of $Z_\T[\bar{L}(n,1)]$ is given by $\langle 0|U_{\gamma(n)}|0\rangle$, which can not be trivial for all invertible $n$. 
In other words the vanishing of all higher central charges is a sufficient condition for the stabilizer of the vacuum state to be $ \Gamma_0(2,\Z_k)$ and hence for theory to admit TBCs.

\section{Proofs of \eqref{toricmass} and \eqref{numberofbinarycodes}}

In this appendix we derive the simplified expressions for the following mass formulas from Section \ref{sum3d}: for $n$ copies of the level $p$ toric code 
\eqref{toricmass} and for $U(1)_2^n$ theory \eqref{numberofbinarycodes}.

Our starting point is the expression for the genus $g$-mass obtained by averaging over genus $g$ Heegaard splittings, 
\bea
 \M_g
 &=&\sum_{h=0}^g \mu(h,g,p)\, \D^h\\
 &=&\sum_{h=0}^g {p^{(g-h)(g-h+1)/2} \over \prod_{i=1}^g (p^i+1)}
\prod_{i=1}^h \frac{p^{g-i+1}-1}{p^i-1} p^{nh}\\
 &=&{1\over \prod_{i=1}^g (p^i+1)}\sum_{h=0}^g p^{(g-h)(g-h+1)/2+nh} 
 \left[
 \begin{matrix}
     g\cr h
 \end{matrix}
 \right]_p, \label{mass3}
\eea 
where the $q$-binomial coefficient is defined in \eqref{binomial}. (In our case $q=p$.)
Using the relation 
\bea
 \left[
 \begin{matrix}
     n\cr m 
 \end{matrix}
 \right]_p = 
  \left[
 \begin{matrix}
     n\cr n-m 
 \end{matrix}
 \right]_p
\eea
we can rewrite the sum in \eqref{mass3} as
\bea
\sum_{h=0}^g  
 p^{h(h+1)/2+n(g-h)}
 \left[
 \begin{matrix}
     g\cr h
 \end{matrix}
 \right]_p 
 &=&p^{gn}\sum_{h=0}^g 
 p^{h(h-1)/2}p^{-(n-1)h}
 \left[
 \begin{matrix}
     g\cr h
 \end{matrix}
 \right]_p
  \\
 &=& 
p^{gn}\prod_{i=0}^{g-1}(1+p^{i-g+1}),\label{mass8}
\eea
where in the second line we made use of the $q$-binomial theorem \cite{kac2002quantum}, 
\bea
\sum_{m=0}^n\, q^{m(m-1)/2}
\left[
 \begin{matrix}
     n\cr m 
 \end{matrix}
 \right]_q t^m 
 = \prod_{i=0}^{n-1} \,(1+t q^i).
\eea
Finally, combining \eqref{mass3} and \eqref{mass8} we obtain 
\bea
 \M_g
=\prod_{i=1}^g  {p^i+p^n\over p^i+1}=p^{gn}\prod_{i=0}^{n-1} {p^i+1\over  p^{i}+p^g}.
\eea 
The last equality may be 
proved by induction on $g$: 
\bea
\prod_{i=1}^g  {p^i+p^n\over p^i+1}&=& {p^g+p^n\over p^g+1}\prod_{i=1}^{g-1}  {p^i+p^n\over p^i+1}\\
&=& {p^g+p^n\over p^g+1} 
p^{(g-1)n}\prod_{i=0}^{n-1} {p^i+1\over  p^{i}+p^{g-1}}\\
&=& {p^g+p^n\over p^g+1} 
p^{(g-1)n} p^{n}\prod_{i=0}^{n-1} {p^i+1\over  p^{i+1}+p^{g}}\\
&=&p^{gn}\prod_{i=0}^{n-1} {p^i+1\over  p^{i}+p^g}.
\eea

The proof of equation \eqref{numberofbinarycodes} is similar.
For finite $g$ and $k=2$ we have from \eqref{Mg}
\bea
M_g =\sum_{h=0}^g \mu(h,g,2)\, \D^h\, \G_h 
=\sum_{h=0}^g {2^{(g-h)(g-h+1)/2}\over \prod_{i=1}^g (2^i+1)} 
\left[
 \begin{matrix}
     g\cr h
 \end{matrix}
 \right]_2
2^{nh\,}2^{-h},
\eea
where we have expressed $\mu(h,g,p)$ as in \eqref{mass3} and inserted  $\D=2^n$ and $\G_h=2^{-h}$ according to \eqref{Gval2}. Continuing as above, we find
\bea
M_g &=& \sum_{h=0}^g {2^{h(h+1)/2}\over \prod_{i=1}^g (2^i+1)} 
\left[
 \begin{matrix}
     g\cr h
 \end{matrix}
 \right]_2
2^{(n-1)(g-h)}\\
&=&
{2^{(n-1)g}\over \prod_{i=1}^g (2^i+1)} \sum_{h=0}^g 2^{h(h-1)/2}
\left[
 \begin{matrix}
     g\cr h
 \end{matrix}
 \right]_2
2^{-(n-2)h}\\
&=& 
{2^{(n-1)g}\over \prod_{i=1}^g (2^i+1)} \prod_{i=0}^{g-1}(1+2^{-(n-2)}2^i)\\
&=&\prod_{i=1}^g {2^i+2^{n-1}\over 2^i+1}\\
&=&2^{g(n-1)}\prod_{i=0}^{n-2} {2^i+1\over 2^i+2^g} \xrightarrow[g\to \infty]{}
\prod_{i=0}^{n-2} (2^i+1)\, .
\eea

\label{identityproof}
%
%

\section{Orthogonal group $O(n,n,\Z_k)$}
\label{orthogonal} 

For general integerl $k$, we define   $\Or(n,n,\Z_k)$ as the
 the symmetry group of matrix 
\bea
\eta=
\left(
\begin{array}{cc}
0 & 1_{n} \\
 1_{n} & 0 \\
\end{array}
\right),
\eea 
with the following important subtelty. For any $k$, including when $k$ is even, 
matrices  $\Or(n,n,\Z_k)$ of the form 
\bea
\left(\begin{array}{cc} 1_n & B \\
0 &1_n \end{array}\right),
\eea
should include only antisymmetric $B$.  Namely, when $k$ is even, we define $\Or(n,n,\Z_k)$  such that the diagonal elements of $B$ vanish to make sure that $\Or(n,n,\Z_k)$ is a symmetry of the corresponding refinement $q$ \eqref{qtoric}.
We also define the subgroup $\Gamma_0(n,n,\Z_k)\subset \Or(n,n,\Z_k)$ of all matrices of the form 
\bea
\left(\begin{array}{cc} A & B \\
0 &D \end{array}\right)\in \Gamma_0(n,n,\Z_k)\subset \Or(n,n,\Z_k).
\eea

Level $k$ toric code always admits TBCs, e.g.~the Dirichlet one for which the Lagrangian subgroup is $\C_0=(\alpha_1,\dots,\alpha_n,0,\dots 0)$ where $\alpha_i\in \Z_k$ are arbitrary.  
The group $\Or(n,n,\Z_k)$ maps Lagrangian subgroups (even self-dual codes) to each other, splitting them into orbits. The subgroup $\Gamma_0$ can be defined as the stabilizer of $\C_0$.
When $k=p$ is a prime, or more generally when $k$ is square-free, there is only one orbit and therefore the total number of codes is given by \cite{Dymarsky:2020qom,Aharony:2023zit}
\bea
&&{|\Or(n,n,\Z_p)|\over |\Gamma_0(n,n,\Z_p)|}=\prod_{i=0}^{n-1}(p^i+1),\\ \nonumber
&&|\Or(n,n,\Z_p)| =2(p^n-1)p^{n(n-1)}\prod_{i=1}^{n-1}(p^{2i}-1),\quad  |\Gamma_0(n,n,\Z_p)|=|\GL(n,\Z_p)| p^{n(n-1)/2}.
\eea

\bibliographystyle{JHEP}
\bibliography{mass}

\end{document}